\documentclass[a4paper,12pt]{article}
\usepackage{psfrag}
\usepackage{afterpage}
\usepackage{multirow}
\usepackage{hhline}
\title{Optimal Spike-Timing Dependent Plasticity for Precise Action Potential Firing}

\author{\small Jean-Pascal Pfister, Taro Toyoizumi\thanks{Current address: Department of Complexity Science and Engineering, Graduate School of Frontier Sciences, The University of Tokyo.}, David Barber\thanks{Current address: IDIAP, Rue du Simplon 4, Case Postale 592, CH-1920 Martigny.}, Wulfram Gerstner \\
\small Laboratory of Computational Neuroscience,\\
\small School of Computer and Communication Sciences\\
\small and Brain-Mind Institute, \\
\small Ecole Polytechnique F\'ed\'erale de Lausanne (EPFL)\\
\small CH-1015 Lausanne\\
\texttt{\small \{jean-pascal.pfister, taro.toyoizumi\}@epfl.ch}\\
\texttt{\small david.barber@idiap.ch, wulfram.gerstner@epfl.ch}
}
\usepackage{graphicx}              

\newcommand \be{\begin{equation}}
\newcommand \ee{\end{equation}}
\newcommand \ba{\begin{eqnarray}}
\newcommand \ea{\end{eqnarray}}
\newcommand \bea{\begin{eqnarray}}
\newcommand \eea{\end{eqnarray}}

\newcommand \nei{i} 

\newcommand \tdes{t^{\rm des}}
\newcommand \tif{t_{\nei}^{f}}
\newcommand \tifm{t_{\nei}^{f-1}}
\newcommand \tpre{t^{\rm pre}}
\newcommand \tjf{t_j^{f}} 
\newcommand \tjk{t_j^{k}} 

\newcommand {\Ex}[1]{\left\langle #1 \right\rangle}

\newcommand \tpost{t^{\rm post}}

\newcommand \Pout{P_{\rm B}}

\newcommand \fmax{F}
\newcommand \tiF{t_{\nei}^{\fmax}}
\newcommand \tio{t_{\nei}^{1}}
\newcommand \tiens{\{\tio,t_{\nei}^{2},\dots,\tiF<t\}}
\newcommand \bx{\mathbf{x}}

\newcommand \Au{A_{\rm u}}
\newcommand \Ac{A_{\rm c}}
\newcommand \Bu{B_{\rm u}}
\newcommand \Bc{B_{\rm c}}
\newcommand \Cu{C_{\rm u}}
\newcommand \Cn{C}
\newcommand \Cc{C_{\rm c}}

\begin{document}
\psfrag{dt}[tc]{$t^{\rm pre}-t^{\rm des}$ [ms]}

\date{}
\maketitle

\begin{abstract}

In timing-based neural codes,
neurons have to  emit action potentials at precise
moments in time. We use a supervised learning
paradigm  to
derive a  synaptic update rule that optimizes
via gradient ascent
the likelihood of postsynaptic firing at
one or several desired firing times.
We find that
the optimal strategy of up- and downregulating synaptic efficacies
can be described by a two-phase learning window similar to that of
Spike-Timing Dependent Plasticity (STDP).
If the presynaptic spike arrives before the desired postsynaptic
spike timing, our optimal learning rule
predicts that the synapse should become potentiated.
The dependence of the potentiation   on spike timing directly reflects
the time course of an excitatory postsynaptic potential.
 The presence and amplitude
of  depression of synaptic efficacies
for reversed spike timing
depends on how constraints are implemented in
the optimization problem.
Two different constraints, i.e.,
 control of postsynaptic rates
or control of temporal locality,
are discussed. 
\end{abstract}

\section{Introduction}

Experimental evidence suggests that precise timing
of spikes is important in several brain systems.
In the barn owl auditory system, for example,
coincidence detecting neurons receive
volleys of temporally precise spikes from both 
ears \cite{Carr90}.
In the electrosensory system of mormyrid electric fish,
medium ganglion cells receive input at precisely
timed delays after electric pulse emission
\cite{Bell97}.
Under the influence  of a common oscillatory
drive as present in the rat hippocampus
or olfactory system, the strength of a constant stimulus is coded in the relative timing of neuronal
action potentials \cite{Hopfield95a,Brody03,Mehta02}.
In humans precise timing of first spikes in
tactile afferents encode touch signals
at the finger tips \cite{Johansson04}.
Similar codes have also been suggested for rapid
visual processing \cite{Thorpe01}, and for the rat's whisker response \cite{Panzeri01b}.

The precise timing of neuronal action potentials also
plays an important role in Spike-Timing Dependent
Plasticity (STDP).
If a presynaptic spike arrives at the synapse
{\em before} the postsynaptic action potential,
the synapse is potentiated;
if the timing is reversed the synapse is depressed \cite{Markram97,Zhang98,Bi98,Bi99,Bi01}.
This biphasic STDP function 
is reminiscent of a temporal contrast
or temporal derivative filter and 
suggests that STDP
is sensitive to the temporal features of a neural
code.
Indeed, theoretical studies have shown that, given 
a biphasic STDP function,
synaptic plasticity
can lead to a stabilization of synaptic weight dynamics
\cite{Kempter99c,Song00,Kempter01,Rossum00,Rubin01}
while the neuron remains sensitive
to temporal structure in the input
\cite{Gerstner96a,Roberts99,Kempter99c,Kistler00,Rao01,Gerstner02b}.

While the relative firing time of pre- and postsynaptic neurons,
and hence temporal aspects of a neural code, 
play a role in STDP, it is, however,
less clear whether STDP is useful to {\em learn} 
a temporal code.
In order to further elucidate the computational function
of STDP, we ask in this paper the following question:
What is the ideal form of a  STDP function
in order to
generate action potentials of the postsynaptic neuron
with high temporal precision?

\section{Model}

\subsection{Coding Paradigm}
\label{sec:paradigm}

In order 
to explain our computational paradigm,
we focus on the example of temporal coding
of human touch stimuli \cite{Johansson04}, 
 but
the same ideas would apply analogously
to the other neuronal systems 
with temporal codes mentioned above
\cite{Carr90,Bell97,Hopfield95a,Brody03,Mehta02,Panzeri01b}.
For a given touch stimulus,
spikes in an ensemble of $N$ tactile afferents
occur in a precise temporal order.
If the same touch stimulus with identical 
 surface properties and force vector is repeated
several times, the relative timing of
action potentials is reliably reproduced
whereas the spike timing in the same
ensemble of afferents is different for other
stimuli \cite{Johansson04}.
In our model, we assume that all input lines, labeled by the index $j$
with $1\le j \le N$
converge onto one or several postsynaptic neurons. 
We think of the postsynaptic neuron as a detector for a given spatio-temporal spike patterns in the input.
The full spike pattern detection paradigm will be used
in Section \ref{scenario-C}. As a preparation and first steps
towards the full coding paradigm we will also consider
the response of a postsynaptic neuron to a single presynaptic spike
 (Section \ref{scenario-A})
or to one given spatio-temporal firing
pattern  (Section \ref{scenario-B}).

\subsection{Neuron Model}

Let us consider a neuron $i$ which is receiving 
input 
from $N$ presynaptic neurons. Let us denote the ensemble of all spikes of neuron $j$ by $x_j = \{t_j^1,\dots,t_j^{N_j}\}$ 
where $\tjk$ denotes the time when neuron $j$ fired
its $k^{\rm th}$ spike. The spatio-temporal spike pattern of all presynaptic neurons $1\leq j\leq N$ will be denoted by boldface $\bx = \{x_1,\dots,x_N\}$.

A presynaptic spike elicited at time $\tjf$  evokes an excitatory postsynaptic potential (EPSP) of amplitude $w_{ij}$ 
and time course $\epsilon(t-\tjf)$. For the sake of simplicity, 
we approximate the EPSP time course by a double exponential 
\begin{equation}
\epsilon(s) = \epsilon_0 \left[\exp\left(-\frac{s}{\tau_m}\right) - \exp\left(-\frac{s}{\tau_s}\right) \right]\Theta(s)
\end{equation}
with a membrane time constant of $\tau_m = 10$ ms and a synaptic time constant of $\tau_s=0.7$ ms 
which yields an EPSP rise time of 2 ms. Here $\Theta(s)$ denotes the Heaviside step function with $\Theta(s) = 1$ for $s>0$ and $\Theta(s) = 0$ else.
We set $\epsilon_0 = 1.3$ mV such that 
a spike at a synapse with $w_{ij}=1$ 
evokes an EPSP with amplitude of approximately 1\,mV. Since the EPSP amplitude is a measure of the strength of a synapse,
we refer to $w_{ij}$ also as the efficacy (or ``weight'') of the synapse between neuron $j$ and $i$.

Let us further suppose  that the  postsynaptic neuron $i$
 receives additional input 
$I(t)$ that could either arise 
from a second group of neurons or 
from intracellular current injection.
We think of the second input as a `teaching' input
that increases the probability that the 
neuron fires at or close to the desired
firing time $\tdes$.
For the sake of simplicity we model
the teaching input as a square current pulse
$I(t) = I_0\Theta(t-\tdes+0.5\Delta T)\Theta(\tdes + 0.5\Delta T - t)$
of amplitude $I_0$ and duration $\Delta T$.
 The effect of the teaching  current on the membrane potential is  
\begin{equation}
u_{\rm teach}(t) = \int_0^{\infty} k(s)I(t-s)ds
\end{equation}
 with $k(s) = k_0\exp(-s/\tau_m)$
where $k_0 = 1/C$ is a constant that is inversely proportional
to the capacitance of the neuronal membrane.

 In the context of the human touch paradigm discussed in section~\ref{sec:paradigm}, the teaching input could represent
some preprocessed visual information
(`object touched by fingers starts to slip now'),
feedback from muscle activity
(`strong counterforce applied now'),
cross-talk from other detector neurons in
the same population
(`your colleagues are active now'),
or unspecific modulatory input due to arousal or reward
(`be aware - something interesting happening now').

The membrane potential of the postsynaptic neuron $\nei$ (Spike Response Model cite{Gerstner02}) is influenced by the EPSPs evoked by all afferent spikes of stimulus $\bx$, the `teaching' signal and 
the refractory effects generated by spikes $\tif$ of the postsynaptic neuron
\be
\label{eq:u}
u_{\nei}(t\vert \bx,y^i_t) =  u_{\rm rest} + \sum_{j=1}^{N} w_{ij}  \sum_{\tjf\in x_j}\epsilon(t-\tjf)
	+ \sum_{\tif\in y^i_t}\eta(t-\tif) + u_{\rm teach}(t)
\ee
where $u_{\rm rest}=-70$ mV is the resting potential, $y^i_t = \tiens$ is the set of postsynaptic spikes that occurred before $t$ and $t_i^F$ always denotes the last postsynaptic spike before $t$. On the right-hand side of Eq.~(\ref{eq:u}), $\eta(s)$ denotes the spike-afterpotential generated by an action potential. We take
\begin{equation} 
\eta(s) = \eta_0\exp\left(-\frac{s}{\tau_m}\right) \, \Theta(s)
\end{equation}
 where $\eta_0<0$  is a  reset parameter 
that describes how much 
the voltage is reset after each spike;
for the relation to  for integrate-and-fire neurons see
\cite{Gerstner02}.
The spikes themselves are not modeled explicitly but reduced to formal firing times. Unless specified otherwise, we take $\eta_0 = -5$ mV.

\begin{figure}[!h]
\psfrag{dw}{$\Delta w^{\Au}$}
\psfrag{isi}[ct]{ISI [ms]}
\psfrag{prob}[cb]{prob. density}
\psfrag{g}[cb]{g [kHz]}
\psfrag{I0}[ct]{$I_0$}
\psfrag{nuout}[cb]{$\nu^{\rm post}$ [Hz]}
\psfrag{pot}[ct]{$u$ [mV]}
\begin{center}     
\begin{tabular}{lll}
{\bf A} & {\bf B} & {\bf C} \\
\includegraphics[width = 0.3\textwidth]{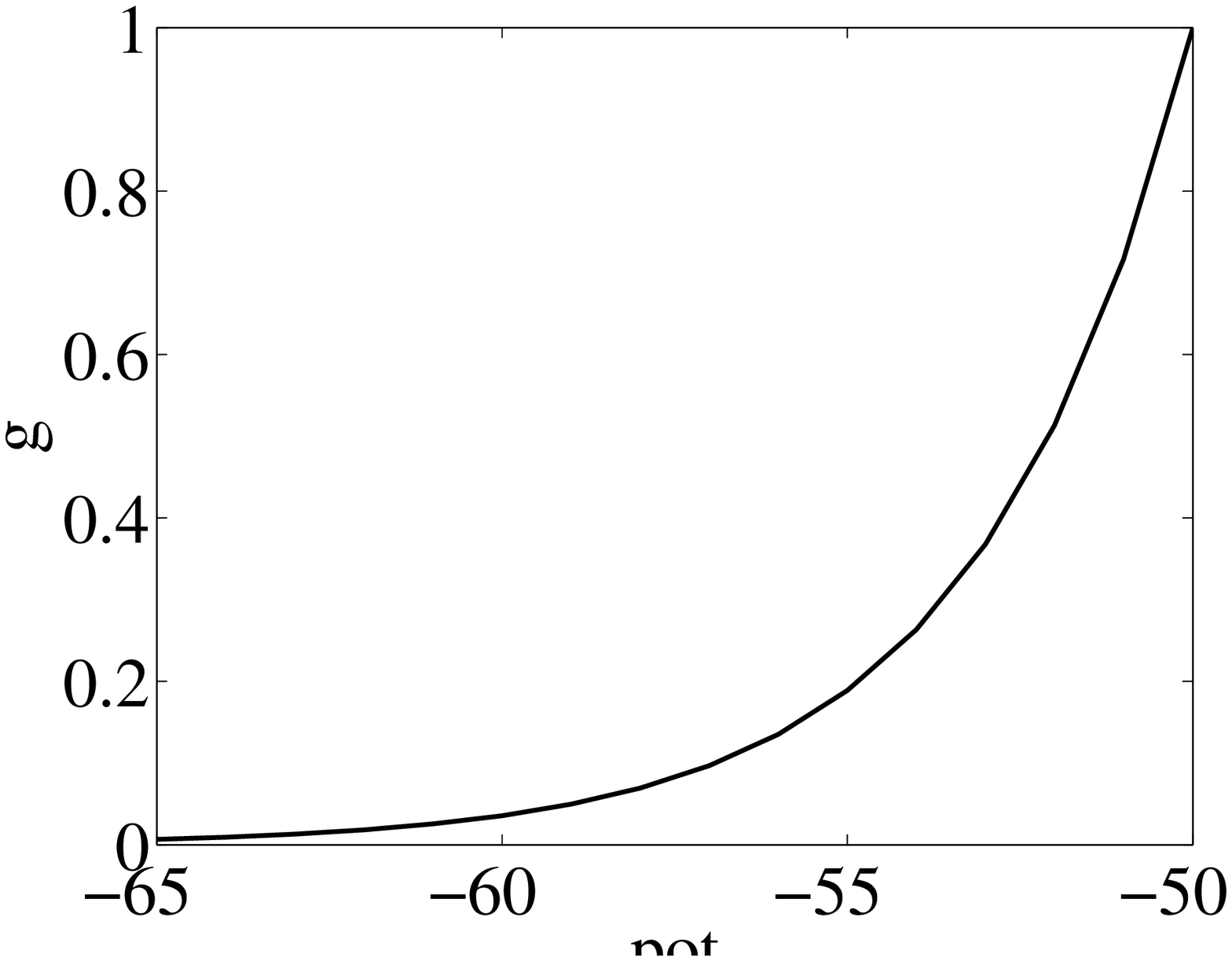} 
& \includegraphics[width = 0.3\textwidth]{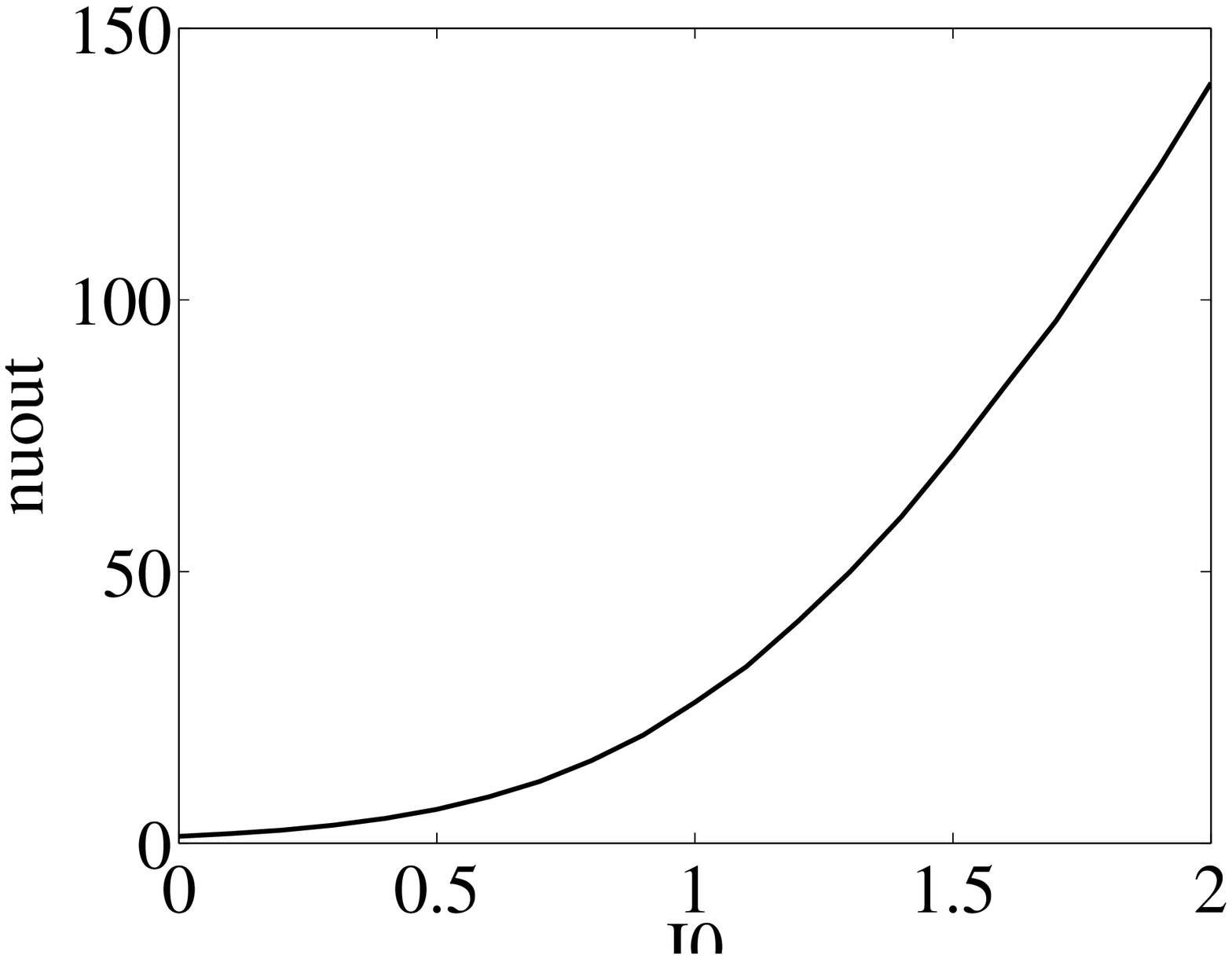}
&\includegraphics[width = 0.3\textwidth]{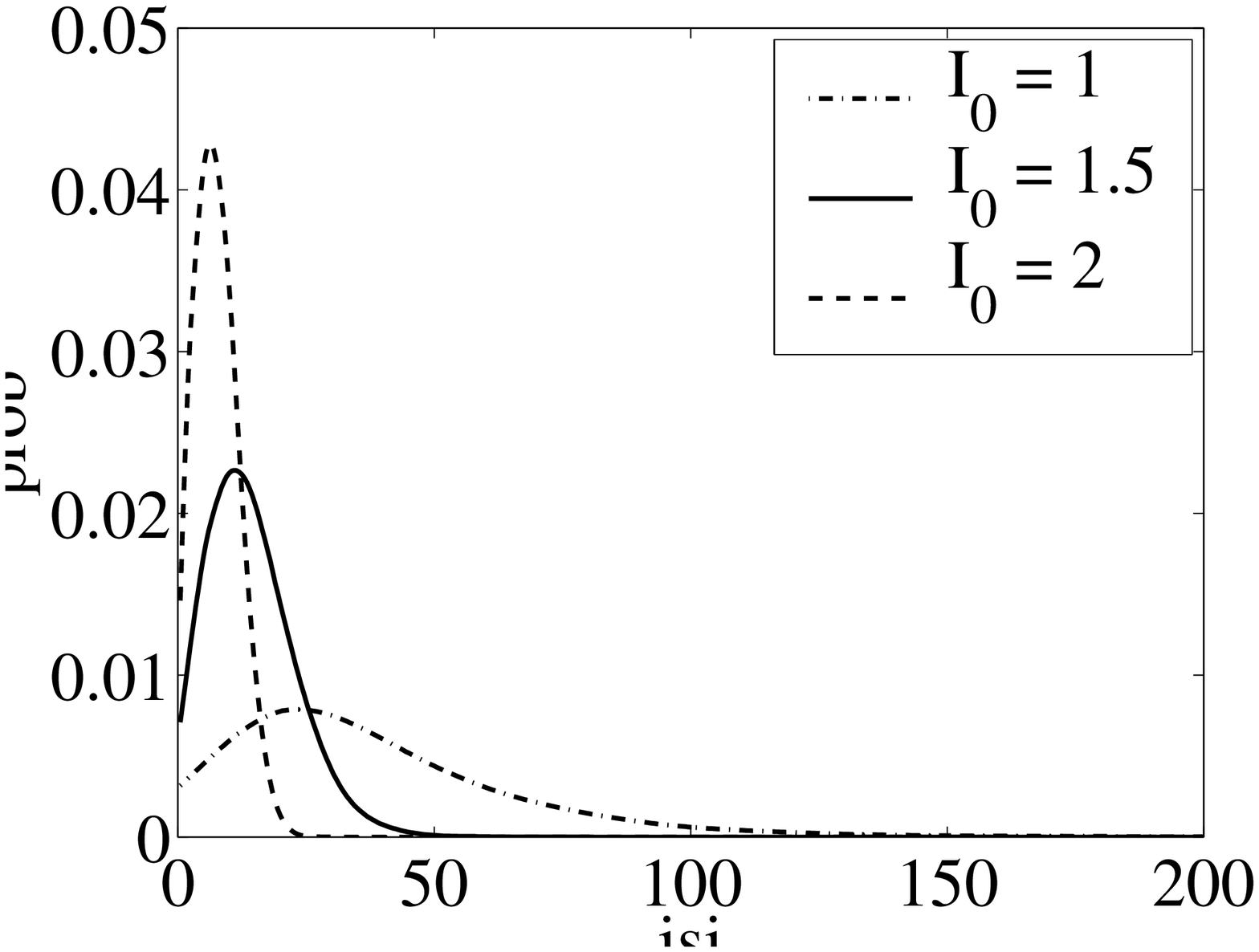}
\end{tabular}
\caption{{\bf A}. Escape rate $g(u) = \rho_0  \exp{\left(u-\vartheta\over \Delta u\right)}$. {\bf B}. Firing rate of the postsynaptic neuron
as a function of the amplitude $I_0$ of a constant stimulation current (arbitrary units).
{\bf C}. Interspike interval (ISI) distribution for different input currents.} 
\label{fig:gain}
\end{center}
\end{figure}

In a deterministic version of the model, output spikes would
be generated whenever the membrane potential $u_i$ reaches 
a threshold $\vartheta$. 
In order to account for intrinsic noise, and also for a small amount of `synaptic noise' generated by 
stochastic spike arrival from additional excitatory and inhibitory 
presynaptic neurons which are not modeled explicitly
we replace the strict threshold by a stochastic one.
More precisely we 
we adopt the following procedure \cite{Gerstner02}. 
Action potentials of the postsynaptic neuron $i$ are generated by a point process with time dependent stochastic intensity $\rho_i(t) = g(u_i(t))$ that depends non-linearly upon the membrane potential $u_i$. Since the membrane potential in turn depends on both the input and the firing history of the postsynaptic neuron, we write:
\be
\label{eq:rho1} 
\rho_i(t\vert \bx,y_t^i) = g(u_i(t\vert \bx,y_t^i)) .
\ee
We take an exponential to describe the escape across a noisy threshold, i.e, $g(u) = \rho_0  \exp{\left(u-\vartheta\over \Delta u\right)}$ where $\vartheta = -50$ mV is the formal threshold, $\Delta u=3$ mV  is the width of the threshold region and therefore tunes the stochasticity of the neuron, and $\rho_0=1$/ms is the stochastic intensity at threshold (see Fig.~\ref{fig:gain}). Other choices of the escape function $g$ are possible with no qualitative change of the results. For $\Delta u \to 0$, the model is identical to the deterministic leaky integrate-and-fire model with synaptic current injection \cite{Gerstner02}.

We note that the stochastic process, defined in Eq.~(\ref{eq:rho1}) is similar to, but different from a Poisson process since the stochastic intensity depends on the set $y_t$ of the previous spikes of the postsynaptic neuron. Thus the neuron model has some `memory' of previous spikes. 

\subsection{Stochastic generative model}

The advantage of the probabilistic framework 
introduced above via the noisy threshold
is that it is possible to describe the probability density\footnote{For the sake of simplicity, we denoted the set of postsynaptic spikes from 0 to $T$ by $y$ instead of $y_T$.} 
$P_i(y\vert \bx)$ of an entire spike train\footnote{Capital $Y$ is the spike train generated by the ensemble (lower case) $y$.} $Y(t) = \sum_{\tif\in y} \delta (t-\tif)$ (see appendix~\ref{sec:PS} for details):
\begin{eqnarray}
P_i(y\vert \bx) &=& \left(\prod_{\tif\in y}\rho_i(\tif\vert \bx,y_{\tif})\right)\exp\left(-\int_0^T\rho_i(s\vert\bx,y_s)ds\right)\nonumber\\
&=& \exp\left(\int_0^T\log(\rho_i(s\vert\bx,y_s))Y(s)-\rho_i(s\vert\bx,y_s)ds\right)\label{eq:Py}
\end{eqnarray}
Thus we have a generative model that allows us to describe 
explicitly the likelihood $P_i(y\vert \bx)$ of emitting a set of spikes $y$ for a given input $\bx$.
Moreover, since  the likelihood  in 
Eq.~(\ref{eq:Py}) is a smooth 
 function of its parameters, 
 it is straightforward to differentiate it  with respect to any variable. Let us differentiate $P_i(y\vert \bx)$ with respect to the synaptic efficacy $w_{ij}$, since this is a quantity that we will use later on:
\begin{equation}
\frac{\partial \log P_i(y\vert \bx)}{\partial w_{ij}} = \int_0^T\frac{\rho'_i(s\vert\bx,y_s)}{\rho_i(s\vert\bx,y_s)}\left[Y(s)-\rho_i(s\vert\bx,y_s)\right]\sum_{\tjf\in x_j}\epsilon(s-\tjf)ds.\label{eq:dlogP}
\end{equation}
where $\rho'_i(s\vert\bx,y_s) = \frac{dg}{du}\vert_{u = u_i(s\vert\bx,y_s)}$. 

In this paper, we propose three different \emph{optimal models} called A, B and C (cf. Table~\ref{tab:scen}). The models differ in the stimulation paradigm and the specific task of the neuron. The common idea behind all three approaches is the notion of optimal performance. Optimality is defined by an objective function $L$ that is directly related to 
the likelihood formula of Eq. \ref{eq:Py}
and that  can be maximized by changes of the synaptic weights. Throughout the paper, this optimization  is done by a standard technique of gradient ascent:
\begin{equation}
\Delta w_{ij} = \alpha\frac{\partial L}{\partial w_{ij}}
\end{equation}
with a learning rate $\alpha$.
Since the three models correspond to three different tasks, they have a slightly different objective function. Therefore, gradient ascent yields slightly different strategies for synaptic update. In the following we start with the simplest model with the aim to illustrate the basic principles that generalize to the more complex models.

\section{Results}

In this section we present
synaptic updates rules derived
by optimizing  the likelihood
of postsynaptic spike firing at
some  desired firing time $\tdes$.
The essence of the argument
is introduced in a particularly
simple scenario, where the neuron
is stimulated by one presynaptic
spike and the neuron is inactive
except at the desired firing
time $\tdes$. This is the raw
scenario
that is further developed
in several different directions.

Firstly, we may ask the question of 
how the postsynaptic spike at the desired time
$\tdes$ is
generated: (i) The spike  could simply be   given
by a supervisor. As always in maximum likelihood
approaches,  we then  optimize
the likelihood that this spike could have
been generated by the neuron model (i.e., the generative model)
given  the known input;
(ii)
the spike could have been  generated by an strong current
pulse of short  duration applied by the supervisor
(teaching input). In this case the a priori 
likelihood that the generative
model fires at or close to the desired firing time is much higher.
The two conceptual paradigms
give slightly different results as 
discussed  in scenario $A$.

Secondly, we may,
in addition to the spike
at the desired time $\tdes$
 allow for
other postsynaptic spikes
generated spontaneously.
The consequences of spontaneous
activity for the STDP function
are discussed in scenario $B$.
Thirdly, instead of imposing
a single postsynaptic spike
at a desired firing time $\tdes$,
we can think of a temporal coding
scheme where the postsynaptic
neuron responds to one  (out of $M$) presynaptic
spike patterns  
with a desired output
spike train containing several spikes
while staying inactive for the
other $M-1$ presynaptic spike patterns.
This corresponds to a pattern classification
task which is the topic of scenario $C$.

Moreover, optimization can be performed
in an unconstrained fashion
or under some constraint.
As we will see in this section,
the specific form of the constraint
influences the results on STDP,
in particular the strength of synaptic
depression for `pre after post' timing.
To emphasize this aspect,
we  discuss two different constraints.
The first constraint is motivated
by the observation that
neurons have a preferred
working point defined by
a typical mean firing rate
that is stabilized by homeostatic
synaptic processes \cite{Turrigiano04}.
Penalizing deviations
from a target firing rate  is the constraint that we will
use in scenario $B$. For very low
target firing rate, the constraint
reduces to the condition of `no activity'
which is the constraint implemented in
scenario $A$.

The second type of constraint is
motivated by the notion of STDP
 itself: changes of synaptic
plasticity should depend on the relative
timing of pre- and postsynaptic spike firing
and not on other factors. If STDP is to be
implemented by some physical or chemical mechanisms
with finite time constants,
we must require  the STDP function to be local
in time, i.e., the amplitude of the STDP function
approaches zero for large time differences.
This is the temporal locality constraint used in scenario
$C$. While the unconstrained optimization
problems are labeled with the subscript $u$ ($\Au$, $\Bu$, $\Cu$), the constrained
problems are marked by the subscript $c$ ($\Ac$, $\Bc$, $\Cc$) (c.f Table~\ref{tab:scen}).

\begin{table}[h]
\begin{center}
\begin{tabular}{|l|l||l|l|}
\hline
\multicolumn{2}{|c||}{Unconstrained scenarios} & \multicolumn{2}{c|}{Constrained scenarios}\\
\hhline{|==#==|}
\multirow{2}{3mm}{$\Au$} & Postsynaptic spike imposed & \multirow{2}{3mm}{$\Ac$} & No activity\\
& $L^{\Au} = \log(\rho(\tdes))$ & & $L^{\Ac} = L^{\Au} - \int_0^{T}\rho(t)dt$ \\
\hline
\multirow{3}{3mm}{$\Bu$} & Postsynaptic spike imposed & \multirow{3}{3mm}{$\Bc$} & \multirow{2}{50mm}{Stabilized activity}\\
& + spontaneous activity & &\\
& $L^{\Bu} = \log(\bar{\rho}(\tdes))$ & & {\scriptsize $L^{\Bc} = L^{\Bu} - \frac{1}{T\sigma^2}\int_0^T(\bar{\rho}(t)-\nu_0)^2dt$} \\
\hline
\multirow{3}{3mm}{$\Cu$} & Postsynaptic spike & \multirow{3}{3mm}{$\Cc$} & Temporal locality\\
& patterns imposed & & constraint\\
&{ \scriptsize $\displaystyle L^{\Cu} = \log\left(\prod_{i}P_i(y^i\vert \bx^i)\prod_{k\neq i}P_i(0\vert \bx^k)^{\frac{\gamma}{M-1}}\right)$} & & {\scriptsize $ L^{\Cc} = L^{\Cu}$, $P_{\Delta\Delta'} = a\delta_{\Delta\Delta'}\left(\Delta-\tilde{T}_0\right)^2$ } \\
\hline
\end{tabular}
\end{center}
\caption{Summary of the optimality criterion $L$ for the three unconstrained scenarios ($\Au$, $\Bu$, $\Cu$) and the three constrained scenarios ($\Ac$, $\Bc$, $\Cc$). The constraint for scenario $C$ is not included in the likelihood function $L^{\Cc}$ itself, but rather in the deconvolution with a matrix $P$ that penalizes quadratically the terms that are non-local in time. See appendix~\ref{sec:deconv} for more details.}
\label{tab:scen}
\end{table}

\subsection{Scenario $A$: ``one postsynaptic spike imposed''}
\label{scenario-A}

Let us start with  a particularly simple model which consists of one presynaptic neuron and one postsynaptic neuron (c.f. Fig.~\ref{fig:ScA}A). 
Let us suppose that the task of the postsynaptic neuron $i$ is to fire a single spike at time $\tdes$ in response to the input which consists of a single presynaptic spike at time $\tpre$, i.e. the input is $x = \{\tpre\}$ and the desired output of the postsynaptic neuron is $y = \{\tdes\}$. 
Since there is only a single pre- and a single postsynaptic
neuron involved, we drop in  this section the indices $j$ and $i$ 
of the two neurons.

\subsubsection{Unconstrained scenario $\Au$: 1 spike at $\tdes$}

In this subsection, we assume that the postsynaptic neuron has not been active in the recent past, i.e. refractory effects are negligible.  In this case, we have $\rho(t\vert \bx,y_t) = \rho(t\vert \bx)$ because of the absence of previous spikes. Moreover, since there is only a single presynaptic spike (i.e. $\bx = \{\tpre\}$), we write $\rho(t\vert\tpre)$ instead of $\rho(t\vert\bx)$.

\begin{figure}[!h]
\psfrag{dw}{$\Delta w^{\Au}$}
\begin{center}     
\begin{tabular}{ll}
{\bf A} & {\bf B} \\
\includegraphics[width = 0.4\textwidth]{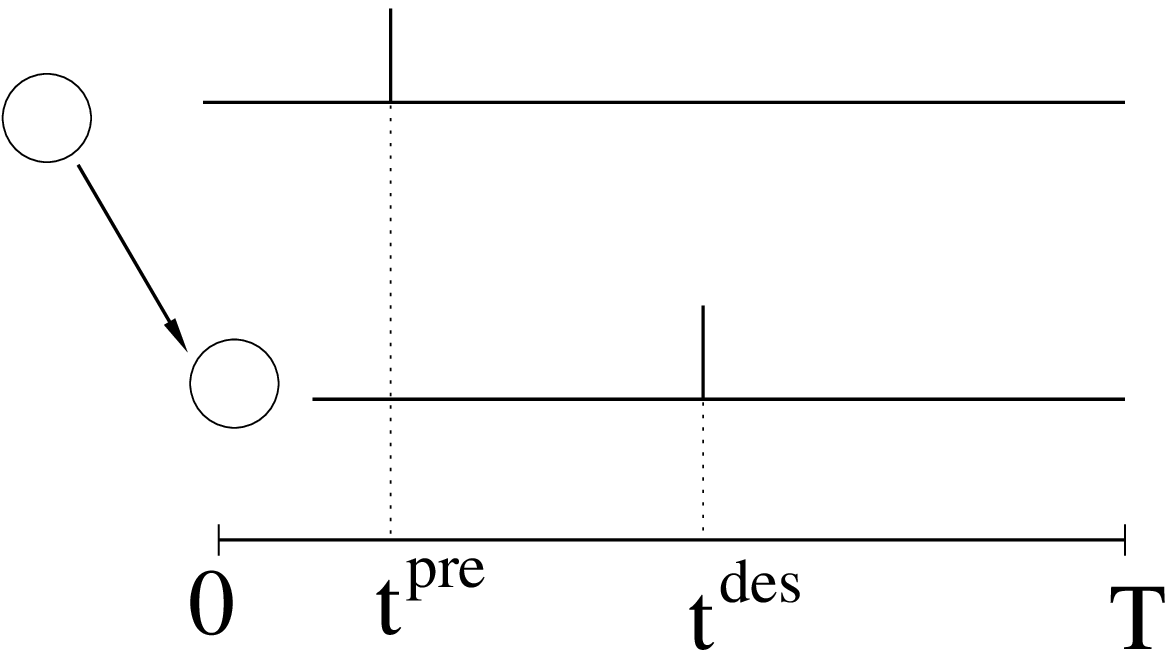} 
& \includegraphics[width = 0.4\textwidth]{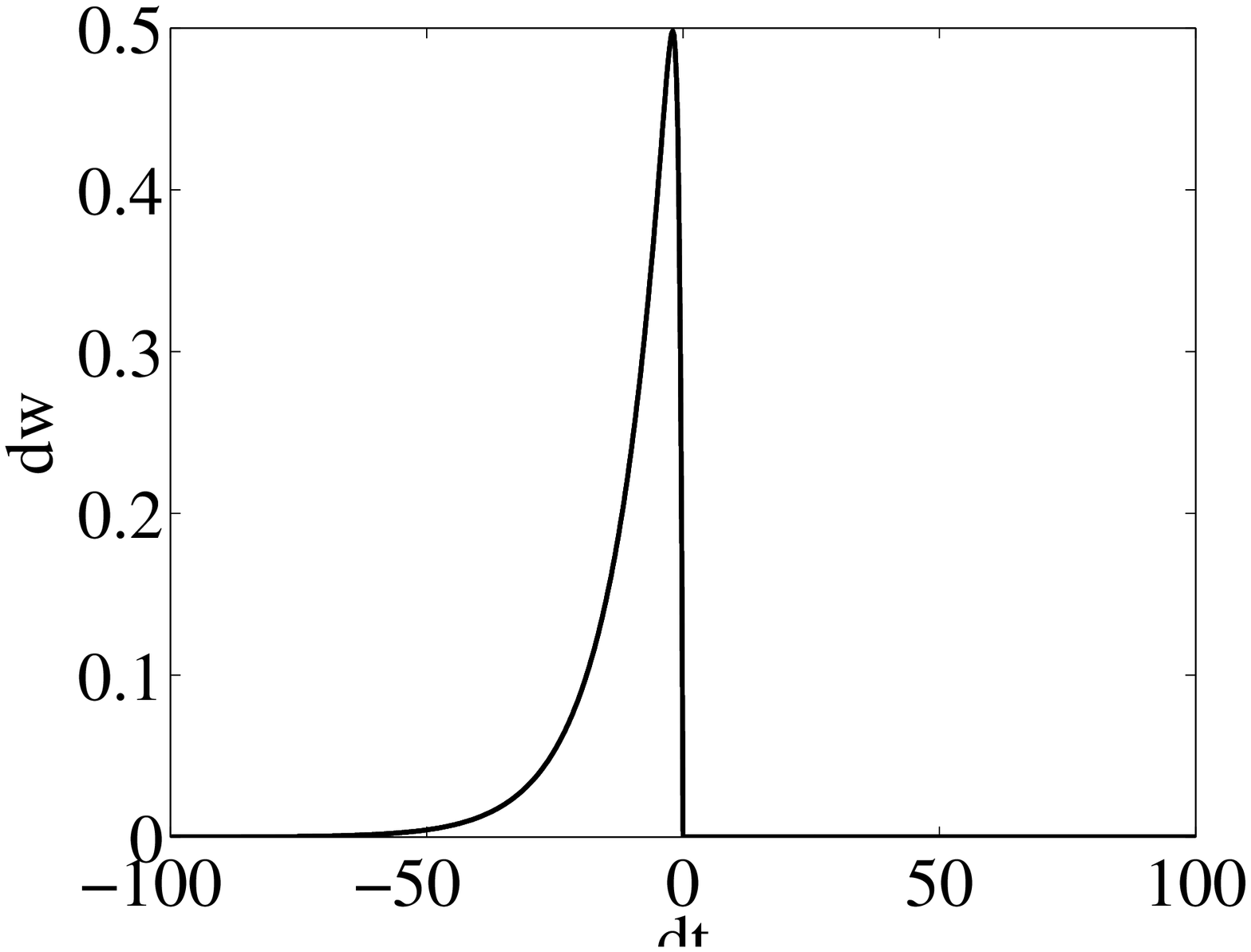}\\
&\\
\end{tabular}
\caption{{\bf A}. Scenario $A$: a single presynaptic neuron connected to a postsynaptic neuron with a synapse of weight $w$. {\bf B}. Optimal weight change given by Eq.~(\ref{eq:dwa0}) for the scenario $\Au$. This weight change is exactly the mirror image of an EPSP.}
\label{fig:ScA}
\end{center}
\end{figure}

Since the task of the postsynaptic neuron is to fire at time $\tdes$, we can define the optimality criterion $L^{\Au}$ as the log-likelihood of the firing intensity at time $\tdes$, i.e. 
\begin{equation}
L^{\Au} = \log\left(\rho(\tdes\vert\tpre)\right)
\end{equation}

The gradient ascent on this function leads to the following STDP function:
\begin{equation}
\Delta w^{\Au} = \frac{\partial L^{\Au}}{\partial w}\nonumber = \frac{\rho'(\tdes\vert\tpre)}{\rho(\tdes\vert\tpre)}\epsilon(\tdes-\tpre)
\label{eq:dwa0}
\end{equation}

where $\rho'(t\vert\tpre) \equiv \frac{dg}{du}\vert_{u=u(t\vert\tpre)}$. Since this optimal weight change $\Delta w^{\Au}$ can be calculated for any presynaptic firing time $\tpre$, we get a STDP function which depends on the time difference $\Delta t = \tpre-\tdes$ (c.f. Fig.~\ref{fig:ScA}B). As we can see directly from Eq.~(\ref{eq:dwa0}), the shape of the potentiation is exactly a mirror image of an EPSP. This result is independent  of the specific choice of the function $g(u)$.

The drawback of this simple model becomes apparent, if the STDP function given by Eq.~\ref{eq:dwa0} is iterated over several repetitions of the experiment. Ideally, it should converge to an optimal solution given by $\Delta w^{\Au} = 0$ in Eq.~(\ref{eq:dwa0}). However, the optimal solution given by $\Delta w^{\Au} = 0$ is problematic: for $\Delta t<0$, the optimal weight tends towards $\infty$ whereas for $\Delta t \geq 0$, there is no unique optimal weight ($\Delta w^{\Au} = 0$, $\forall w$). The reason of this problem is, of course, that the model describes only potentiation and includes no mechanisms for depression.

\subsubsection{Constrained scenario $\Ac$: ``no other spikes than at $\tdes$''}

\begin{figure}[!h]
\psfrag{dw}{$\Delta w^{\Ac}$}
\begin{center}                  
\begin{tabular}{ll}
{\bf A} \hspace{1cm} without teaching & {\bf B} \hspace{1cm} with teaching\\
\includegraphics[width = 0.4\textwidth]{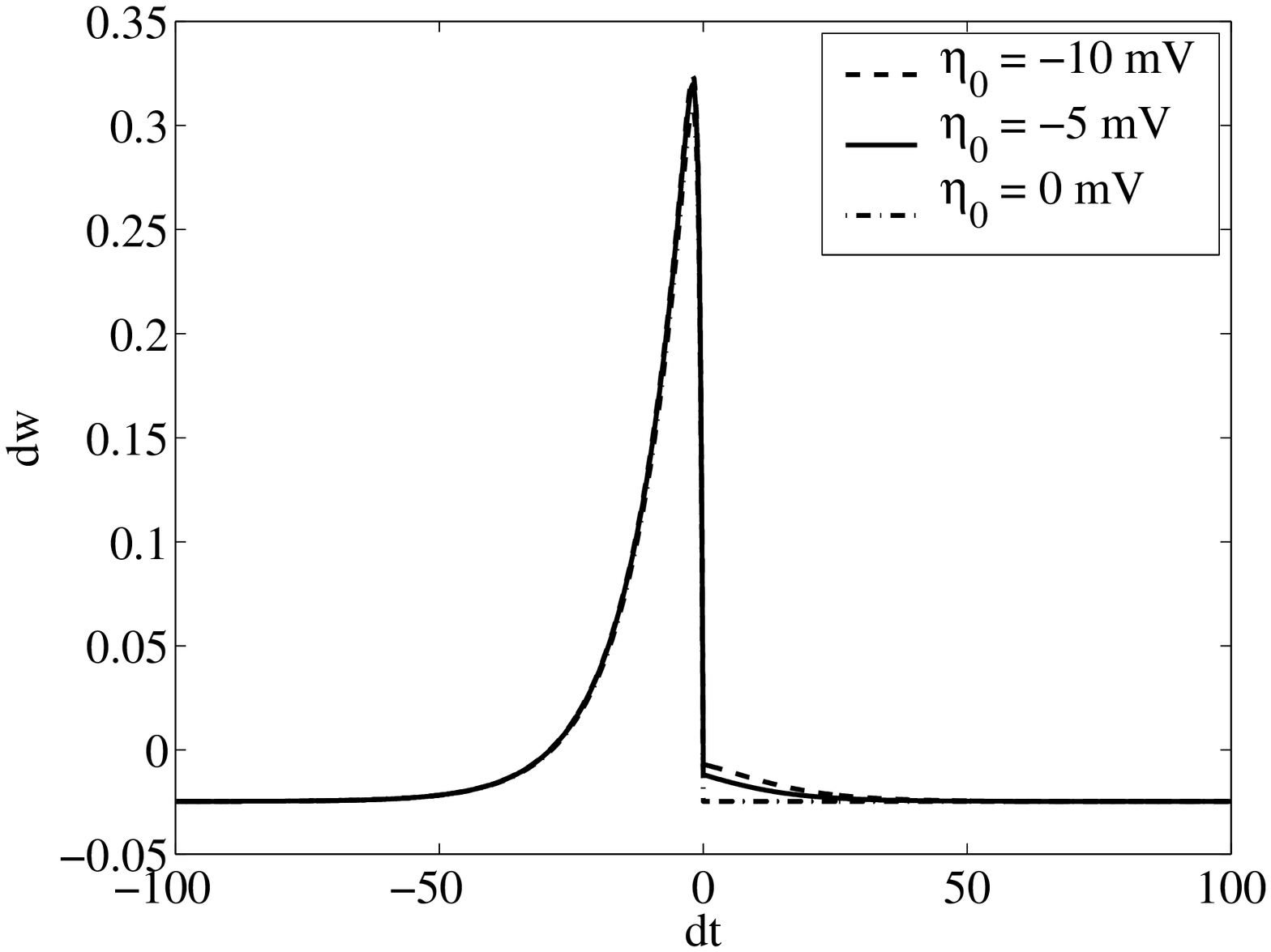} &
\includegraphics[width = 0.4\textwidth]{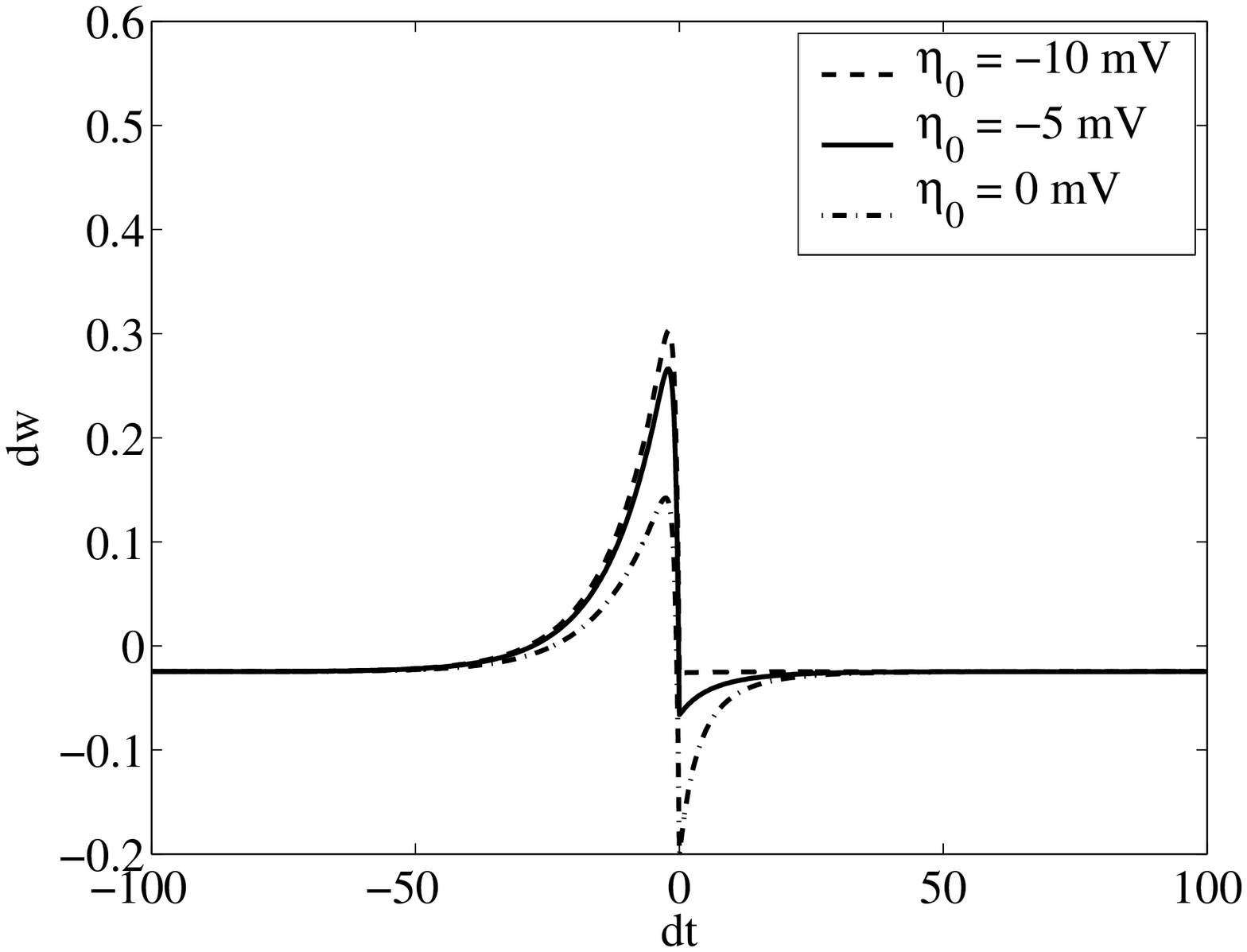}
\end{tabular}
\caption{Optimal weight adaptation for scenario $\Ac$ given by Eq.~(\ref{eq:dlogP}) in the case of a teaching signal ({\bf A}) and in the absence of a teaching signal ({\bf B}). The weight change in the post-before-pre region is governed by the spike afterpotential $u_{AP}(t) = \eta(t)+u_{\rm teach}(t)$. The duration of the teaching input is $\Delta T = 1$ ms. The amplitude of the current $I_0$ is chosen so that $\max_t u_{\rm teach}(t) = 5$ mV. $u_{\rm rest}$ is chosen such that the spontaneous firing rate $g(u_{\rm rest})$ matches the desired firing rate $1/T$, i.e. $u_{\rm rest} = \Delta u\log\frac{\tau_0}{T}+\theta \simeq -60$ mV. The weight strength is $w = 1$.}
\label{fig:A1}
\vspace{0.2cm}
\end{center}
\end{figure}

In order to get some insight of where the depression could come from, let us consider a small modification of the previous model. In addition to the fact that the neuron has to fire at time $\tdes$, let us 
 suppose that it should not fire anywhere else. This condition can be implemented by an application of Eq.~(\ref{eq:Py}) to the case of a single input spike $x = \{\tpre\}$ and a single output spike $y = \{\tdes\}$.
In terms of notation we set $P(y\vert x) = P(\tdes\vert\tpre)$
and similarly $\rho(s\vert x,y) = \rho(s|\tpre,\tdes)$
and use  Eq.~(\ref{eq:Py}) to find:
\begin{equation}
\label{A-c1}
P(\tdes\vert\tpre) = \rho(\tdes\vert \tpre) \exp\left[ -\int_0^T\rho(s\vert \tpre,\tdes)ds \right] \, .
\end{equation}
Note that for $s\leq\tdes$, the firing intensity does not depend on $\tdes$, hence $\rho(s\vert\tpre,\tdes) = \rho(s\vert\tpre)$ for $s\leq\tdes$.
We define the objective function $L^{\Ac}$ as the log-likelihood of generating a single output spike at time $\tdes$, given a single input spike at $\tpre$. Hence, with Eq.~(\ref{A-c1}):
\begin{eqnarray}
L^{\Ac} &=& \log(P(\tdes \vert \tpre)) \nonumber \\
&=&  \log(\rho(\tdes\vert \tpre)) -\int_0^T\rho(s\vert \tpre,\tdes)ds \label{eq:LA1}
\end{eqnarray}
and the gradient ascent $\Delta w^{\Ac} = \partial L^{\Ac}/\partial w$ 
rule yields
\begin{equation}
\label{w-Ac}
\Delta w^{\Ac} = \frac{\rho'(\tdes\vert\tpre)}{\rho(\tdes\vert\tpre)}\epsilon(\tdes-\tpre) - \int_0^T\rho'(s\vert \tpre,\tdes)\epsilon(s-\tpre)ds\label{eq:dwa1}
\end{equation}
Since we have a single postsynaptic spike at $\tdes$,
Eq.~(\ref{eq:dwa1}) can directly be plotted as a STDP function.
In Fig.~\ref{fig:A1} we distinguish two different cases.
In Fig.~\ref{fig:A1}A we optimize the likelihood  $L^{\Ac}$ 
in the absence of  any teaching input.
To understand this scenario we may imagine that
a postsynaptic spike has occurred spontaneously at the 
desired firing time $\tdes$. Applying the appropriate
weight update calculated from Eq. (\ref{w-Ac}) will make
such a timing more likely the next time the presynaptic stimulus
is repeated. The reset amplitude $\eta_0$ has only a small influence.

In Fig.~\ref{fig:A1}B we consider a case where firing
of the postsynaptic spike at the appropriate time 
was made highly likely by a teaching input of duration
$\Delta T = 1$ ms centered around the desired firing $\tdes$.
The form of the STDP function depends on the 
amount $\eta_0$ of the reset.
If there is no reset $\eta_0=0$,
the STDP function shows strong synaptic depression
of synapses that become active {\em after}
the postsynaptic spike. This is due 
to the fact that the teaching input causes
an  increase of the membrane potential that
decays back to rest with the 
membrane time constant $\tau_m$.
Hence the window of synaptic depression 
is also exponential with the same time constant.
Qualitatively the same is true, if we include
a weak reset. The from of the depression window
remains the same, but its amplitude is reduced.
Only for strong reset to or below resting potential
the effect inverses. A weak reset
is standard in applications of integrate-and-fire
models to in vivo data and is one
of the possibilities to explain the high
coefficient of variation of neuronal spike trains
in vivo 
\cite{Bugmann97,Troyer97}.

A further property of the STDP functions in 
 Fig.~\ref{fig:A1} is a negative offset
for $|\tpre-\tdes| \to \infty$.
The amplitude of the offset 
 can be calculated for $w\simeq 0$ and $\Delta t>$, i.e. $\Delta w_0 \simeq -\rho'(u_{\rm rest}) \int_0^{\infty}\epsilon(s)ds$.  This offset is due to the fact that we do not want spikes at other times than $\tdes$. As a result, the optimal weight $w^{\star}$ (i.e. solution of $\Delta w^{\Au}$), should be as negative as possible ($w^{\star} \rightarrow -\infty$ or $w^{\star} \rightarrow w^{\rm min}$ in the presence of a lower bound) for $\Delta t > 0$ or $\Delta t \ll 0$.

\subsection{Scenario $B$: ``spontaneous activity''}
\label{scenario-B}
The constraint in  Scenario $A_c$  of having strictly no other postsynaptic spikes than the one at time $\tdes$ 
may seem artificial. 
Moreover, it is this constraint which leads to 
the negative  offset of the STDP function
discussed at the end of the previous paragraph.
In order to relax the constraint of ``no spiking'', we allow 
in scenario $B$ for a reasonable spontaneous activity.
As above, we start with an unconstrained scenario $B_u$
before we turn to the constrained scenario $B_c$.

\subsubsection{Unconstrained scenario $\Bu$: maximize the firing rate at $\tdes$ }

Let us start with the simplest model which includes  spontaneous activity. 
Scenario $\Bu$ is the analog of the model $\Au$, but with two differences. 

First, we include spontaneous activity in the model. Since $\rho(t\vert \bx, y_t)$ depends on the spiking history for any given trial, we have to define a quantity which is independent of the specific realizations $y$ of the postsynaptic spike train. 

\begin{figure}[!h]
\psfrag{dw}{$\Delta w^{\Bu}$}
\begin{center}
\begin{tabular}{ll} 
{\bf A} & {\bf B}\\
\includegraphics[width = 0.35\textwidth]{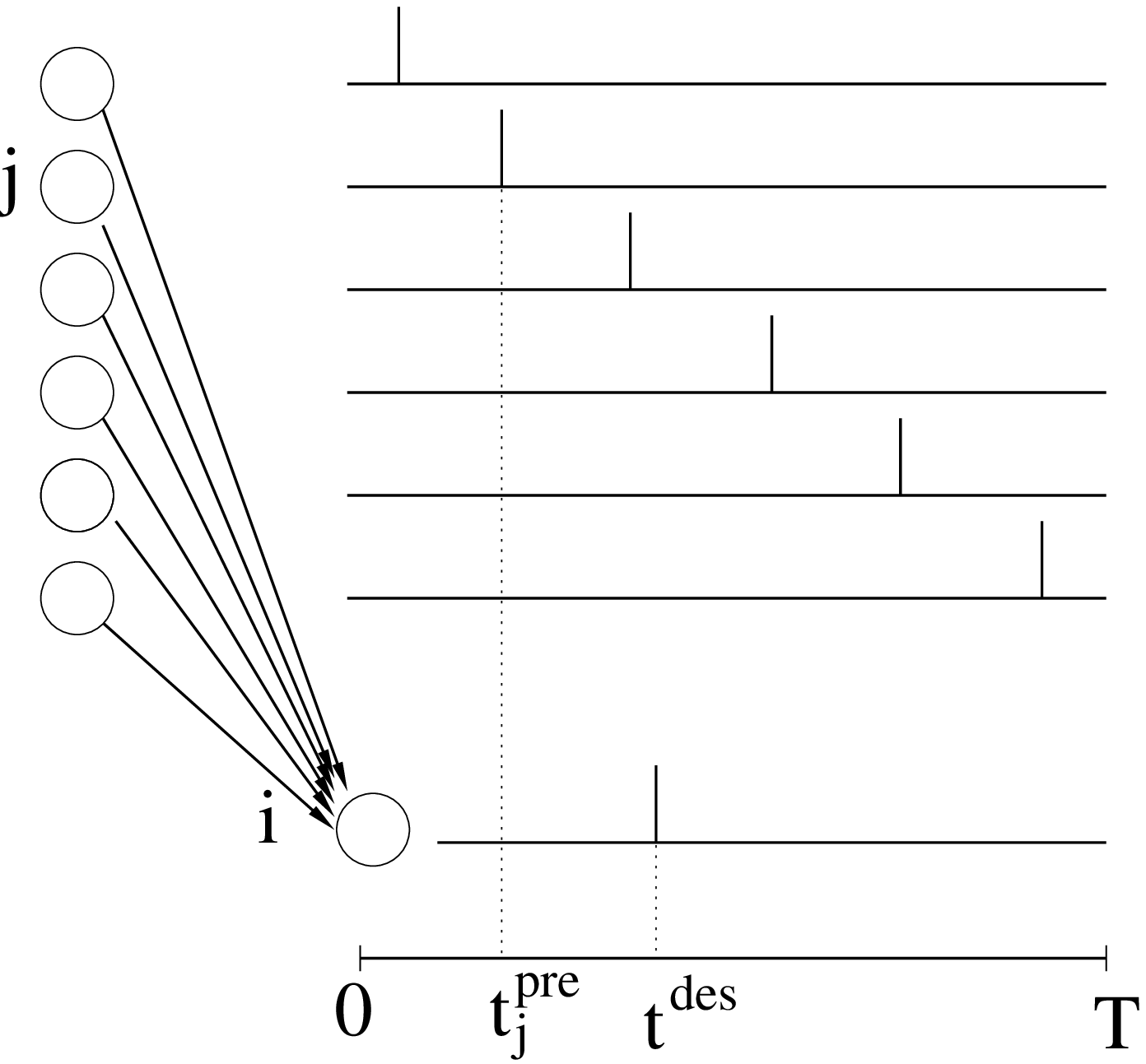}&
\includegraphics[width = 0.55\textwidth]{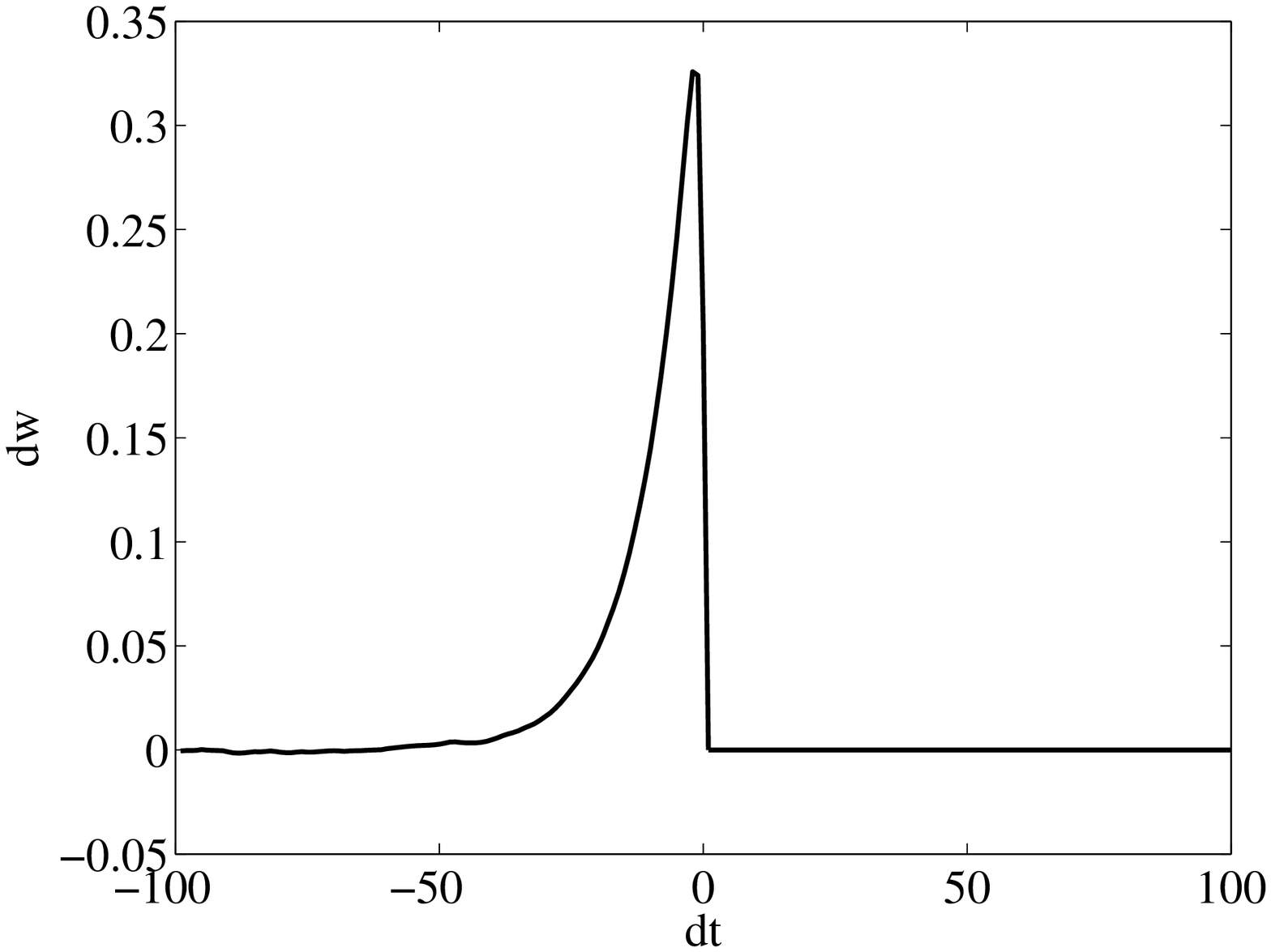}\\
\end{tabular}
\caption{Scenario $B$. {\bf A}. $N = 200$ presynaptic neurons are firing one after the other at time $t_j = j\delta t$ with $\delta t = $ 1 ms. {\bf B}. The optimal STDP function of scenario $\Bu$.}
\label{fig:ScB}
\end{center}
\end{figure}

Secondly, instead of considering only one presynaptic neuron, we consider $N = 200$ presynaptic neuron, each  of them emitting a single spike at time $t_j = j\delta t$, where $\delta t = 1$ ms (see Fig.~\ref{fig:ScB}A). The input pattern will therefore be described by the set of delayed spikes $\bx = \{x_j = \{t_j\},j=1,\dots,N\}$. As long as we consider only a single spatio-temporal spike pattern in the input, it is always possible to relabel neurons appropriately so that neuron $j+1$ fires after neuron $j$.

Let us define the instantaneous firing rate $\bar{\rho}(t)$  that can be calculated by averaging $\rho(t\vert y_t)$ over all realizations of postsynaptic spike trains: 
\begin{equation}
\label{eq:brho}
\bar{\rho}(t\vert \bx) = \Ex{\rho(t\vert \bx, y_t)}_{y_t\vert \bx}.
\end{equation}
Here the notation $\Ex{\cdot}_{y_t\vert\bx}$ means taking the average over all possible configuration of postsynaptic spikes up to $t$ for a given input $\bx$. In analogy to a Poisson process, a specific spike train with firing times $y_t = \tiens$ is generated with probability $P(y_t\vert\bx)$ given by Eq.~(\ref{eq:Py}). Hence, the average $\Ex{\cdot}_{y_t\vert\bx}$ of Eq.~(\ref{eq:brho}) can be written as follows (see appendix~\ref{sec:numerical} for numerical evaluation of $\bar{\rho}(t)$):
\begin{equation}
\label{eq:brho2}
\bar{\rho}(t\vert\bx) = \sum_{\fmax = 0}^{\infty}\frac{1}{\fmax!}\int_0^t\dots\int_0^t\rho(t\vert \bx,y_t)P(y_t\vert\bx)d\tiF\dots\tio.
\end{equation}
Analogously to the model $\Au$, we can define the quality criterion as the log-likelihood $L^{\Bu}$ of firing at the desired time $\tdes$: 
\begin{equation}
L^{\Bu} = \log(\bar{\rho}(\tdes\vert\bx)).
\label{eq:barrho}
\end{equation}
Thus the optimal weight adaptation of synapse $j$ is given by
\begin{equation}
\Delta w_j^{\Bu} = \frac{\partial\bar{\rho}(\tdes\vert\bx)/\partial w_j}{\bar{\rho}(\tdes\vert\bx)}
\end{equation}
where $\frac{\partial \bar{\rho}(t\vert\bx)}{\partial w_j}$ is given by
\begin{equation}
\frac{\partial \bar{\rho}(t\vert\bx)}{\partial w_j} = \bar{\rho}'(t\vert\bx)\epsilon(t-t_j) + \Ex{\rho(t\vert\bx,y_t)\frac{\partial}{\partial w_j}\log P(y\vert\bx)}_{y\vert\bx},
\end{equation}
$\frac{\partial}{\partial w_j}\log P(y)$ is given by Eq.~(\ref{eq:dlogP}) and $\bar{\rho}'(t) = \Ex{\frac{dg}{du}\big\vert_{u \equiv u(t)}}_{y}$.

Figure~\ref{fig:ScB}B shows that, for our standard set of parameters, the differences to scenario $\Au$ are negligible.

Figure~\ref{fig:B1}A depicts 
the STDP function 
 for various values of the parameter $\Delta u$ at a higher postsynaptic firing rate. 
We can see a small undershoot in the pre-before-post region. 
The presence of this small undershoot can be understood as follows: enhancing a synapse of a presynaptic neuron that fires too early would induce a postsynaptic spike that arrives before the desired firing time and therefore, 
because of refractoriness, would prevent the generation of a spike at the desired time. 
The depth of this undershoot decreases with the stochasticity of the neuron and increases with the amplitude of the refractory period (if $\eta_0 = 0$, there is no undershoot). 
In fact, correlations between pre- and postsynaptic firing reflect the shape of an EPSP in the high-noise regime, whereas they show a trough for low noise \cite{Poliakov97,Gerstner01}. Our theory shows that the pre-before-post region of the optimal plasticity function is a mirror image of these correlations.

\subsubsection{Constrained scenario $\Bc$: ``firing rate close to $\nu_0$''}

\begin{figure}[!h]

\psfrag{dt}[tc]{$\tpre -\tdes$ [ms]}
\begin{center}                  
\begin{tabular}{ll} 
{\bf A} & {\bf B}\\
{\psfrag{dw}{$\Delta w^{\Bu}$}
\includegraphics[width = 0.4\textwidth]{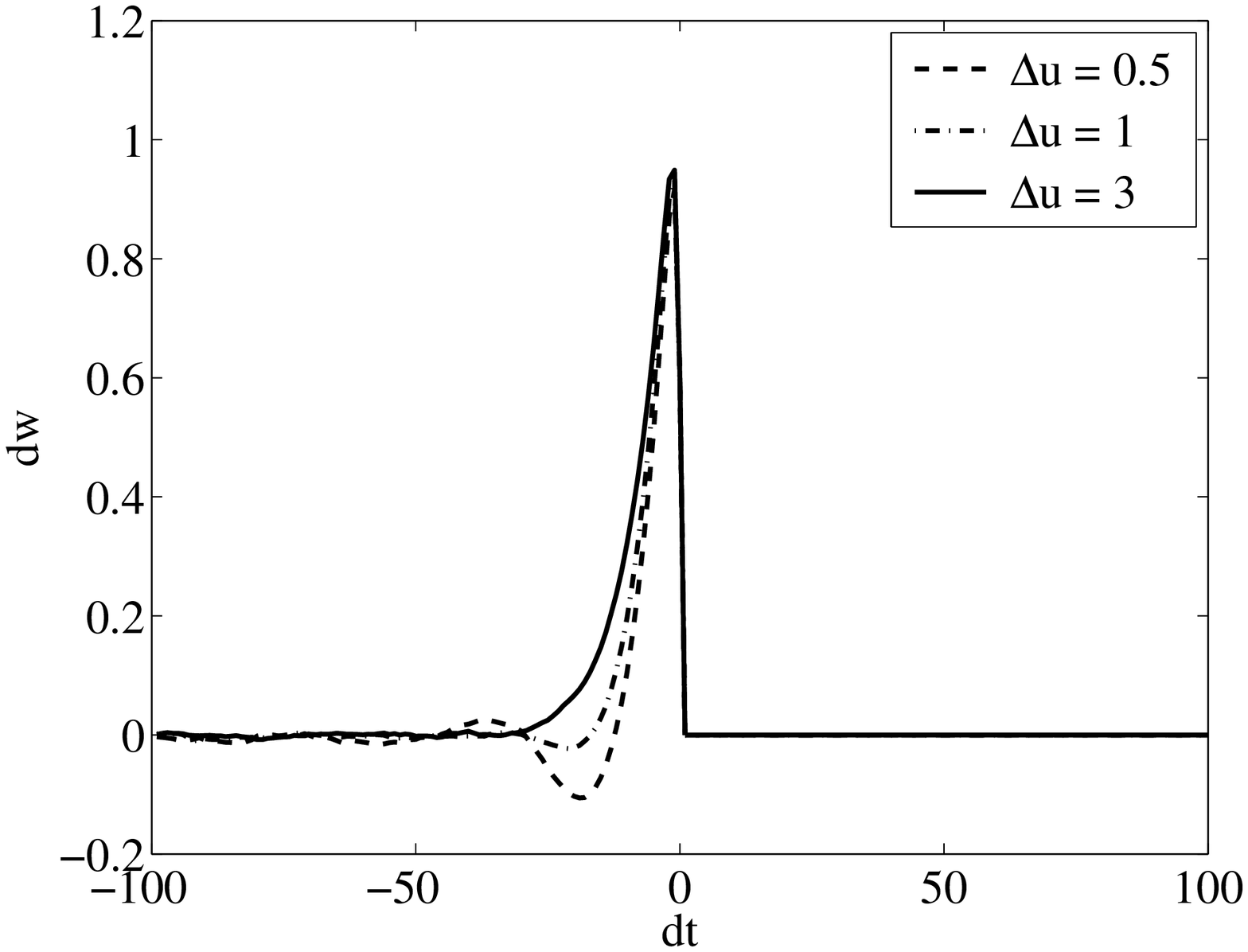}}&
{\psfrag{dw}{$\Delta w^{\Bc}$}
\includegraphics[width = 0.4\textwidth]{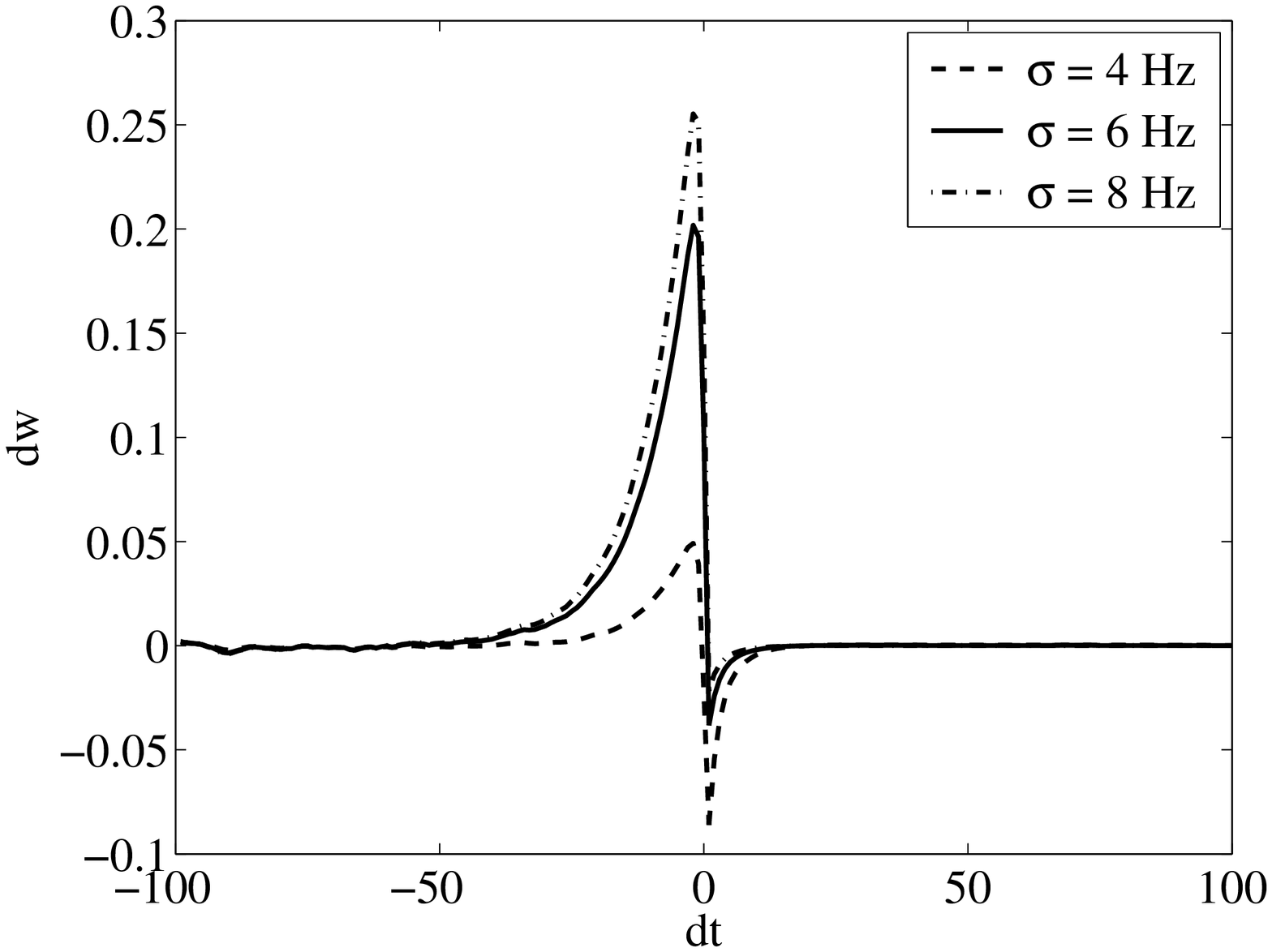}}\\
\end{tabular}
\caption{{\bf A}. The optimal STDP functions of scenario $\Bu$ for different level of stochasticity described by the parameter $\Delta u$. 
The standard value ($\Delta u = 3$ mV) is given by the solid line,
decreased noise ($\Delta u = 1$ mV and $\Delta u = 0.5$ mV)
are indicated by dot-dashed and dashed lines respectively .
 In the  low-noise regime, enhancing a synapse which fires slightly too early can prevent the firing at the desired firing time $\tdes$ due to refractoriness. To increase the firing rate at $\tdes$
it is thence advantageous to decrease the firing probability 
some time before $\tdes$. Methods:
For each value of $\Delta u$, the initial weight $w_0$ are set such that the spontaneous firing rate is $\bar{\rho} = 30$Hz. The amplitude of the STDP function is multiplied by $\Delta u$ in order to make them comparable. Reset: $\eta_0 = -5 $ mV. {\bf B}. Scenario $\Bc$. 
Optimal STDP function for scenario $\Bc$ given by Eq.~(\ref{eq:dwB1}) for a teaching signal of duration $\Delta T = 1$ ms. 
The maximal increase of the membrane potential 
after 1 ms of stimulation with the teaching input
  is $\max_t u_{\rm teach}(t) = 5$ mV.  
Synaptic efficacies $w_{ij}$ are initialized such that $u_0 = -60$ mV which gives a spontaneous rate of $\bar{\rho} = \nu_0 = 5$ Hz. Standard noise level: $\Delta u = 3$ mV.}
\label{fig:B1}
\end{center}
\end{figure}

In analogy to  model $\Ac$ 
we now introduce a constraint.
 Instead of imposing strictly no spikes at times $t\neq \tdes$, we can relax the condition and minimize deviations of the instantaneous firing rate $\bar{\rho}(t\vert\bx,\tdes)$ from a reference firing rate $\nu_0$. This can be done by introducing into Eq.~(\ref{eq:barrho}) a penalty term $\Pout$ given by
\begin{equation}
\Pout =  \exp\left(-\frac{1}{T}\int_0^T\frac{(\bar{\rho}(t\vert\bx,\tdes)-\nu_0)^2}{2\sigma^2}dt\right).
\label{eq:Pout}
\end{equation}
For small $\sigma$, deviations from the reference rate yields a large penalty. For $\sigma\rightarrow\infty$, the penalty term has no influence. The optimality criterion is a combination of a high firing rate $\bar{\rho}$ at the desired time under the constraint of small deviations from the reference rate $\nu_0$. If we impose the penalty as a multiplicative factor and take
as before  the logarithm, we get:
\begin{equation}
L^{\Bc} = \log\left(\bar{\rho}(\tdes\vert\bx)\Pout\right)
\label{eq:LB1}
\end{equation}
Hence the optimal weight adaptation is given by
\begin{equation}
\Delta w_j^{\Bc} = \frac{\partial \bar{\rho}(\tdes\vert\bx)/\partial w_j}{\bar{\rho}(\tdes\vert\bx)}-\frac{1}{T\sigma^2}\int_0^T(\bar{\rho}(t\vert\bx,\tdes)-\nu_0)\frac{\partial}{\partial w_j}\bar{\rho}(t\vert\bx,\tdes)dt.  \label{eq:dwB1}
\end{equation}
Since in scenario $B$ each presynaptic neuron $j$ fires exactly once at time $t_j = j\delta t$ and the postsynaptic neuron is trained to fire at time $\tdes$, we can interpret the weight adaptation $\Delta w_j^{\Bc}$ of Eq.~(\ref{eq:dwB1}) as a STDP function $\Delta w^{\Bc}$ which depends on the time difference $\Delta t = \tpre-\tdes$. Fig.~\ref{fig:B1} shows this STDP function for different values of the free parameter $\sigma$ of Eq.~(\ref{eq:Pout}). The higher the standard deviation $\sigma$, the less effective is the penalty term. In the limit of $\sigma \rightarrow \infty$, the penalty term can be ignored and the situation is identical to that of  scenario $\Bu$.

\subsection{Scenario $C$: ``pattern detection''}
\label{scenario-C}

\subsubsection{Unconstrained scenario $\Cu$: ``spike pattern imposed''}

\begin{figure}[!h]
\psfrag{x1}{$\bx^1$}
\psfrag{y1}{$y^1$}
\begin{center}
\includegraphics[width = 0.8\textwidth]{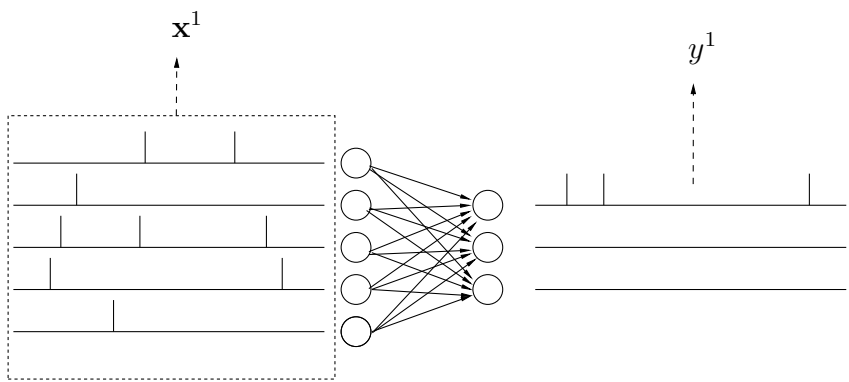} 
\caption{Scenario $C$. $N$ presynaptic neurons are fully connected to $M$ postsynaptic neurons. Each postsynaptic neuron is trained to respond to a specific input pattern and not respond to $M-1$ other patterns as described by the objective function of Eq.~(\ref{eq:LC0}).}
\label{fig:ScC}
\end{center}
\end{figure}

This last scenario is a generalization of the scenario $\Ac$. Instead of restricting the study to a single pre- and postsynaptic neuron, we consider $N$ presynaptic neurons and $M$ postsynaptic neurons (see Fig.~\ref{fig:ScC}). The idea is to construct $M$ independent \emph{detector neurons}. Each detector neuron $i = 1,\dots, M$, should respond best to a specific prototype stimulus, say $\bx^i$, by producing a desired spike train $y^i$,  but should not respond to other stimuli, i.e. $y^i = 0$,  $\forall \bx^k$, $k\neq i$ (see Fig.~\ref{fig:ScC}). The aim is to find a set of synaptic weights that maximize the probability that neuron $i$ produces $y^i$ when $\bx^i$ is presented and produces no output when $\bx^k$, $k\neq i$ is presented. Let the likelihood function $L^{\Cu}$ be
\begin{equation}
L^{\Cu} = \log\left(\prod_{i = 1}^{M}P_i(y^i\vert \bx^i)\prod_{k=1,k\neq i}^{M}P_i(0\vert \bx^k)^{\frac{\gamma}{M-1}}\right) \label{eq:LC0}
\end{equation}

where $P_i(y^i\vert \bx^i)$ (c.f Eq.~(\ref{eq:Py}))  is the probability that neuron $i$ produces the spike train $y^i$ when the stimulus $\bx^i$ is presented. The parameter $\gamma$ characterizes the relative importance of the patterns that should not be learned compared to those that should be learned. We get 
\begin{equation}
L^{\Cu} = \sum_{i=1}^{M}\log(P_i(y^i\vert \bx^i)) - \gamma\Ex{\log(P_i(0\vert \bx^k))}_{\bx^k\neq\bx^i}
\end{equation}

where the notation $\Ex{\cdot}_{\bx^k\neq\bx^i} \equiv \frac{1}{M-1}\sum_{k\neq i}^{M}$ means taking the average over all patterns other than $\bx^i$. The optimal weight adaptation yields
\begin{equation}
\Delta w_{ij}^{\Cn} = \frac{\partial}{\partial w_{ij}}\log(P_i(y^i\vert \bx^i)) -\gamma\Ex{\frac{\partial}{\partial w_{ij}}\log(P_i(0\vert \bx^k))}_{\bx^k\neq\bx^i}
\label{eq:dwC0}
\end{equation}

The learning rule of Eq.~(\ref{eq:dwC0}) gives the optimal weight change for each synapse and can be evaluated after presentation of all pre- and postsynaptic spike patterns, i.e. it is a ``batch'' update rule. Since each pre- and postsynaptic neuron emit many spikes in the interval $[0,T]$, we can not directly interpret the result of Eq.~(\ref{eq:dwC0}) as a function of the time difference $\Delta t = \tpre-\tpost$ as we did in scenario $A$ or $B$.

\begin{figure}[!h]
\psfrag{gp}[bc]{pattern to be learned}
\psfrag{bp}[bc]{pattern not to be learned}
\psfrag{xi}{$\bx^i$}
\psfrag{yi}{$y^i$}
\psfrag{xk}{$\bx^k$}
\psfrag{yk}{$y^k$}
\psfrag{ri}{$\rho_i$}
\psfrag{rk}{$\rho_k$}
\psfrag{tim}[tc]{time [ms]}
\begin{center}
\includegraphics[width = 0.8\textwidth]{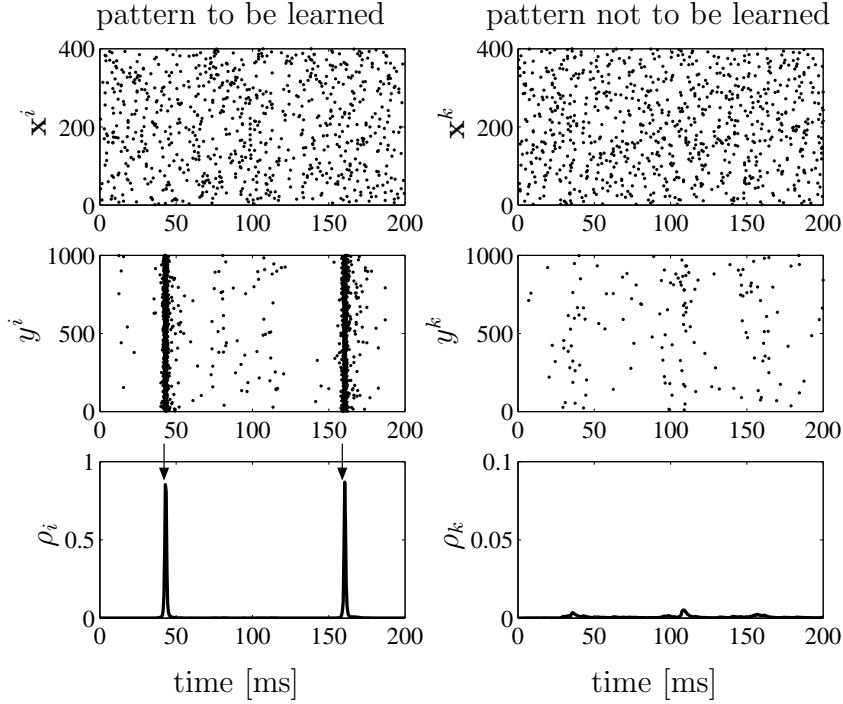}
\caption{Pattern detection after learning. {\bf Top}. The \emph{left} raster plot represents the input pattern the $i^{\rm th}$ neuron has to be sensitive to. Each line corresponds to one of the $N = 400$ presynaptic neurons. Each dot represents an action potential. The \emph{right} figure represents one of the patterns the $i^{\rm th}$ neuron should not respond to. {\bf Middle}. The \emph{left} raster plot corresponds to 1000 repetitions of the output of neuron $i$ when the corresponding pattern $\bx^i$ is presented. The \emph{right} plot is the response of neuron $i$ to one of the pattern it should not respond to. {\bf Bottom}. The \emph{left} graph represents the probability density of firing when pattern $\bx^i$ is presented. This plot can be seen as the PSTH of the middle graph. Arrows indicate  the supervised timing neuron $i$ learned. The \emph{right} graph describes the probability density of firing when pattern $\bx^k$ is presented. Note the different scales of vertical axis.}
\label{fig:ScCa}
\end{center}
\end{figure}

\begin{figure}[!h]
\begin{center}
\psfrag{dt}[tc]{$t^{\rm pre}-t^{\rm post}$ [ms]}
\begin{tabular}{ll}
{\bf A} & {\bf B}\\
{\psfrag{dw}{$\Delta W^{\Cu}$}
\includegraphics[width = 0.4\textwidth]{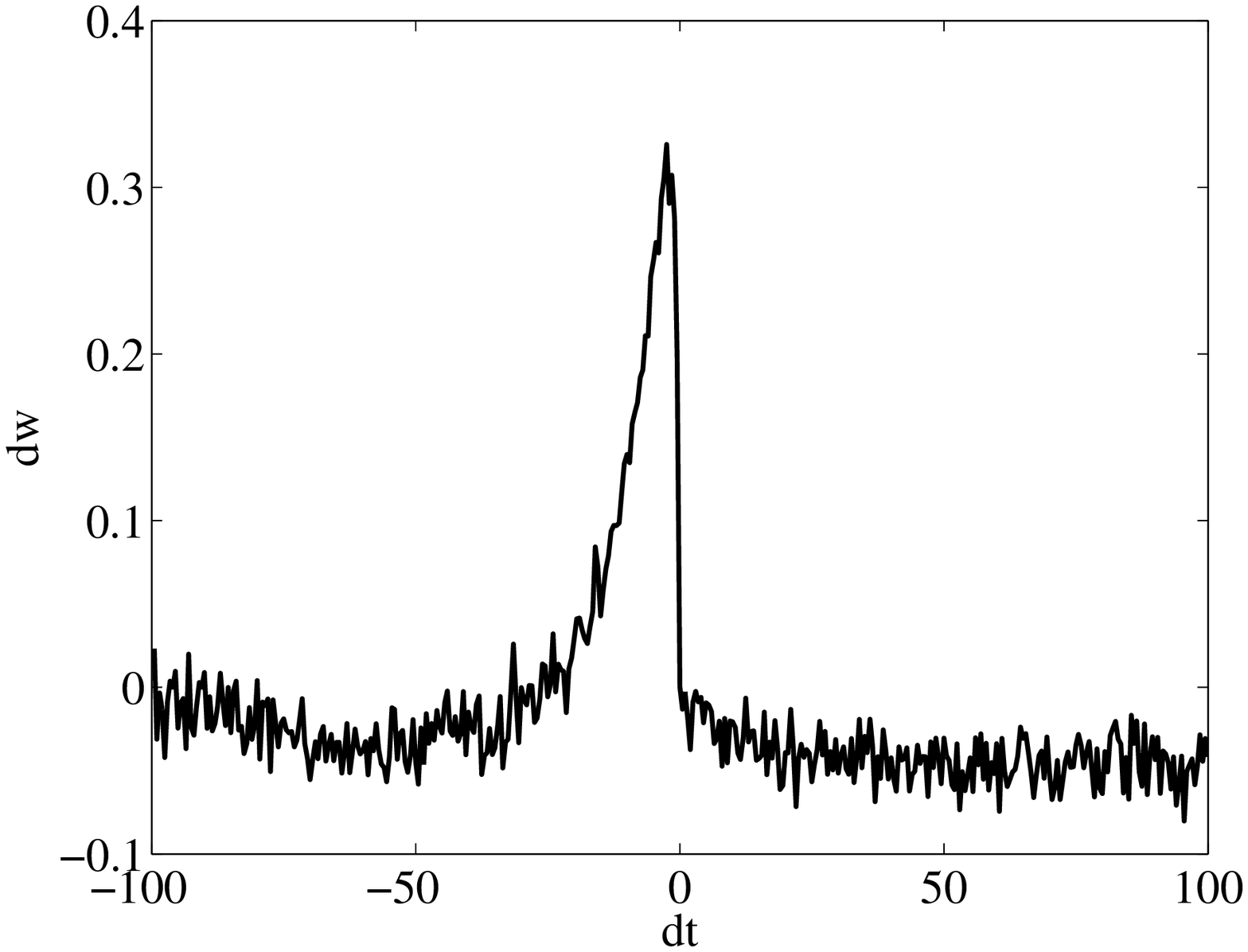}} &
{\psfrag{dw}{$\Delta  W^{\Cc}$}
\includegraphics[width = 0.4\textwidth]{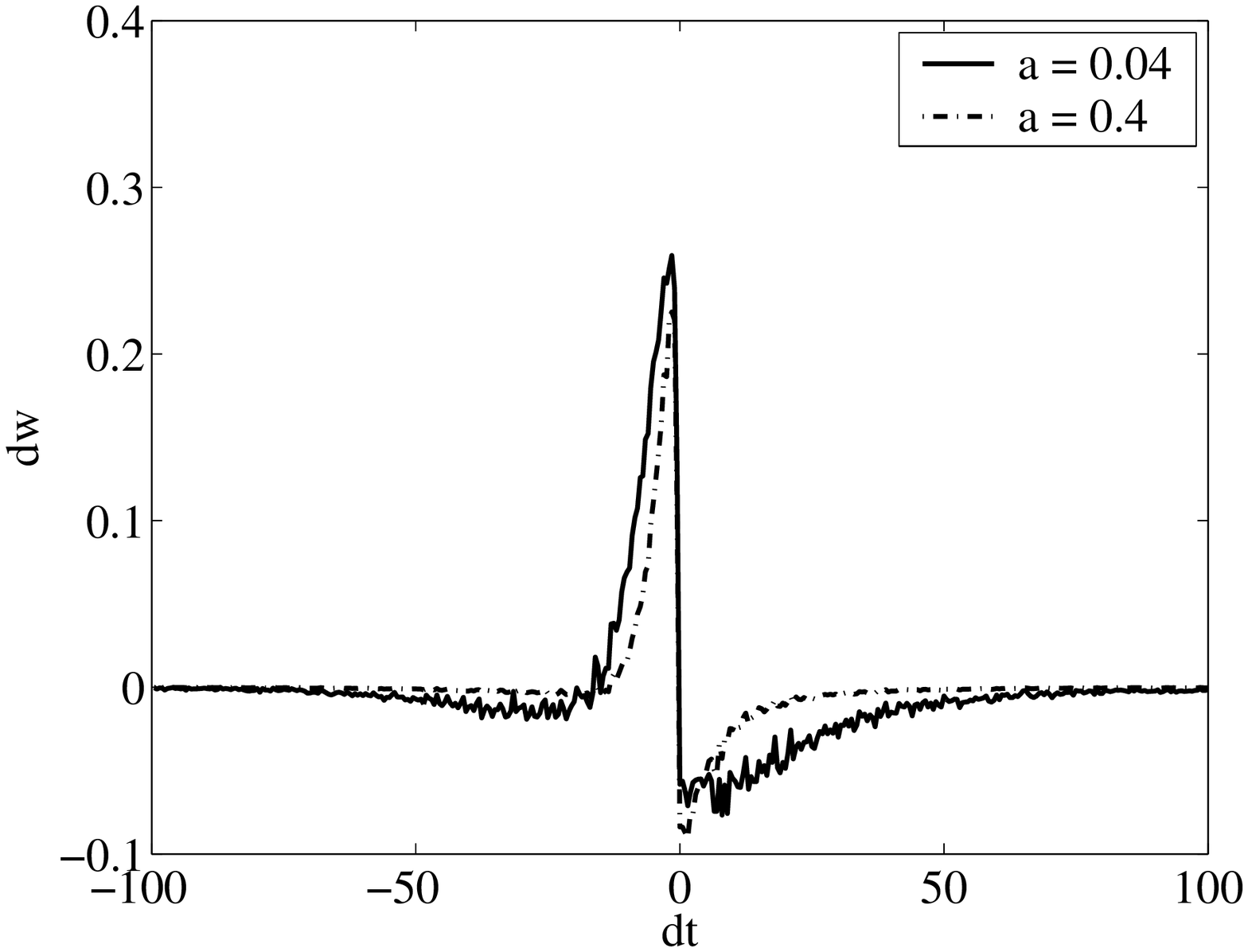}}
\end{tabular}
\caption{{\bf A}. Optimal weight change for scenario $\Cu$. In this case,  no locality constraint is imposed and the result is similar to the STDP function of scenario $\Ac$ (with $\eta_0 = 0$ and $u_{\rm teach}(t) = 0$)  represented on Fig.~\ref{fig:A1}.  {\bf B}. Optimal weight change for scenario $\Cc$ as a function of the locality constraint characterized by $a$. The stronger the importance of the locality constraint, the narrower is the spike-spike interaction. For {\bf A} and {\bf B}: $M = 20$, $\eta_0 = -5$ mV. The initial weights $w_{ij}$ are chose so that the spontaneous firing rate matches the imposed firing rate.}
\label{fig:C1}
\end{center}
\end{figure}

Ideally, we would like to write the total weight change of the optimal rule given by Eq.~(\ref{eq:dwC0}) as a sum of contributions 
\begin{equation}\label{sumrule}
\Delta w_{ij}^{\Cn} = \sum_{
\begin{array}{l}
\tpre\in x_j^i\\
\tpost\in y^i
\end{array}
}\Delta W^{\Cu}(\tpre-\tpost),
\end{equation}
where $\Delta W^{\Cu}(\tpre-\tpost)$ is a STDP function and the summation runs over all pairs of pre- and postsynaptic spikes.
The number of pairs of pre- and postsynaptic spikes
with a given time shift is given by the correlation function
which is best defined in discrete time.
We assume time steps of duration   $\delta t = 0.5$ ms.
Since the correlation will depend on the presynaptic neuron $j$
{\em and} the postsynaptic neuron $i$ under consideration,
we introduce a new index $k=N(i-1)+j$.
We define the correlation in discrete time 
by its matrix elements
 $C_{k\Delta}$ that describe  the correlation between the presynaptic spike train $X_j^i(t)$ and the postsynaptic spike train $Y^i(t-T_0+\Delta\delta t)$.
For example, $C_{3\Delta} = 7$ implies that 7 spike pairs 
of presynaptic neuron $j=3$ with postsynaptic neuron $i=1$
have a relative time shift of $T_0 -\Delta\delta t$.
With this definition, we can rewrite
Eq. (\ref{sumrule}) in vector notation
(see appendix~\ref{sec:deconvpair} for more details):
\begin{equation}
\Delta\mathbf{w}^{\Cn} \stackrel{!}{=}  C\Delta\mathbf{W}^{\Cu}
\label{eq:dwCDW}
\end{equation}
where $\Delta \mathbf{w}^{\Cn} = (\Delta w_{11}^{\Cn},\dots,\Delta w_{1N}^{\Cn},\Delta w_{21}^{\Cn},\dots,\Delta w_{MN}^{\Cn})^T$ is the vector containing all the optimal weight change given by  Eq.~(\ref{eq:dwC0}) and $\Delta \mathbf{W}^{\Cu} $ is the vector containing the discretized STDP function with components $\Delta W_{\Delta}^{\Cu} = \Delta W^{\Cu}(-T_0+\Delta\delta t)$ for $1\leq\Delta\leq 2\tilde{T}_0$ with $\tilde{T}_0 = T_0/\delta$. In particular, the center of th STDP function (i.e. $\tpre = \tpost$) corresponds to the index $\Delta = \tilde{T}_0$. The symbol $\stackrel{!}{=}$ expresses the fact that we want to find $\Delta\mathbf{W}^{\Cu}$ such that $\Delta\mathbf{w}^{\Cn}$ is as close as possible to $C\Delta\mathbf{W}^{\Cu}$. By taking the pseudo-inverse $C^+ = (C^TC)^{-1}C^T$ of $C$, we can invert Eq.~(\ref{eq:dwCDW}) and get
\begin{equation}
\Delta \mathbf{W}^{\Cu} = {C^{+} \Delta \mathbf{w}^{\Cn}}
\label{eq:Cinverse}
\end{equation}
The resulting STDP function is plotted in Fig.~\ref{fig:C1}A. As it was the case for the scenario $\Au$, the STDP function exhibits a negative offset. In addition to the fact the postsynaptic neuron $i$ should not fire at other times than the ones given by $y^i$, it should also not fire whenever pattern $\bx^k$, $k\neq i$ is presented. 
The presence of the negative offset is due to those two factors.

\subsubsection{Constrained scenario $\Cc$: ``temporal locality''}
In the previous paragraph, we obtained a STDP function with a negative offset. This negative offset does not seem realistic because it implies that the STDP function is not localized in time. 
In order to  impose temporal locality (finite memory span of the learning
rule) we  modify Eq.~(\ref{eq:Cinverse}) in the following way (see appendix~\ref{sec:deconvlocal} for more details):
\begin{equation}
\Delta \mathbf{W}^{\Cc} = (C^TC + P)^{-1}C^T\Delta \mathbf{w}^{\Cn}
\end{equation}
where $P$ is a diagonal matrix which penalizes non-local terms. 
In this paper, we take a quadratic suppression of terms that are 
non-local in time. With respect to a postsynaptic spike at $\tpost$,
the penalty term is proportional to $(t-\tpost)^2$. In matric notation, and using our convention that the postsynaptic spike corresponds to $\Delta = \tilde{T}_0$, we have:
\begin{equation}
P_{\Delta\Delta'} = a\delta_{\Delta\Delta'}\left(\Delta-\tilde{T}_0\right)^2
\end{equation}
The resulting STDP functions for different values of $a$ are plotted in Fig.~\ref{fig:C1}B. The higher the parameter $a$, the more non-local terms are penalized, the narrower is the STDP function. 

\begin{figure}[!h]
\psfrag{dwreal}[tc]{$\Delta {\bf w}^{\rm opt}$}
\psfrag{dwest}[bc]{$\Delta {\bf w}^{\rm rec}$}
\begin{center}
\begin{tabular}{lll}
{\bf A} \hspace{0.07\textwidth} $a = 0$ & {\bf B} \hspace{0.07\textwidth} $a = 0.04$ & {\bf C} \hspace{0.07\textwidth} $a = 0.4$\\
\includegraphics[width = 0.3\textwidth]{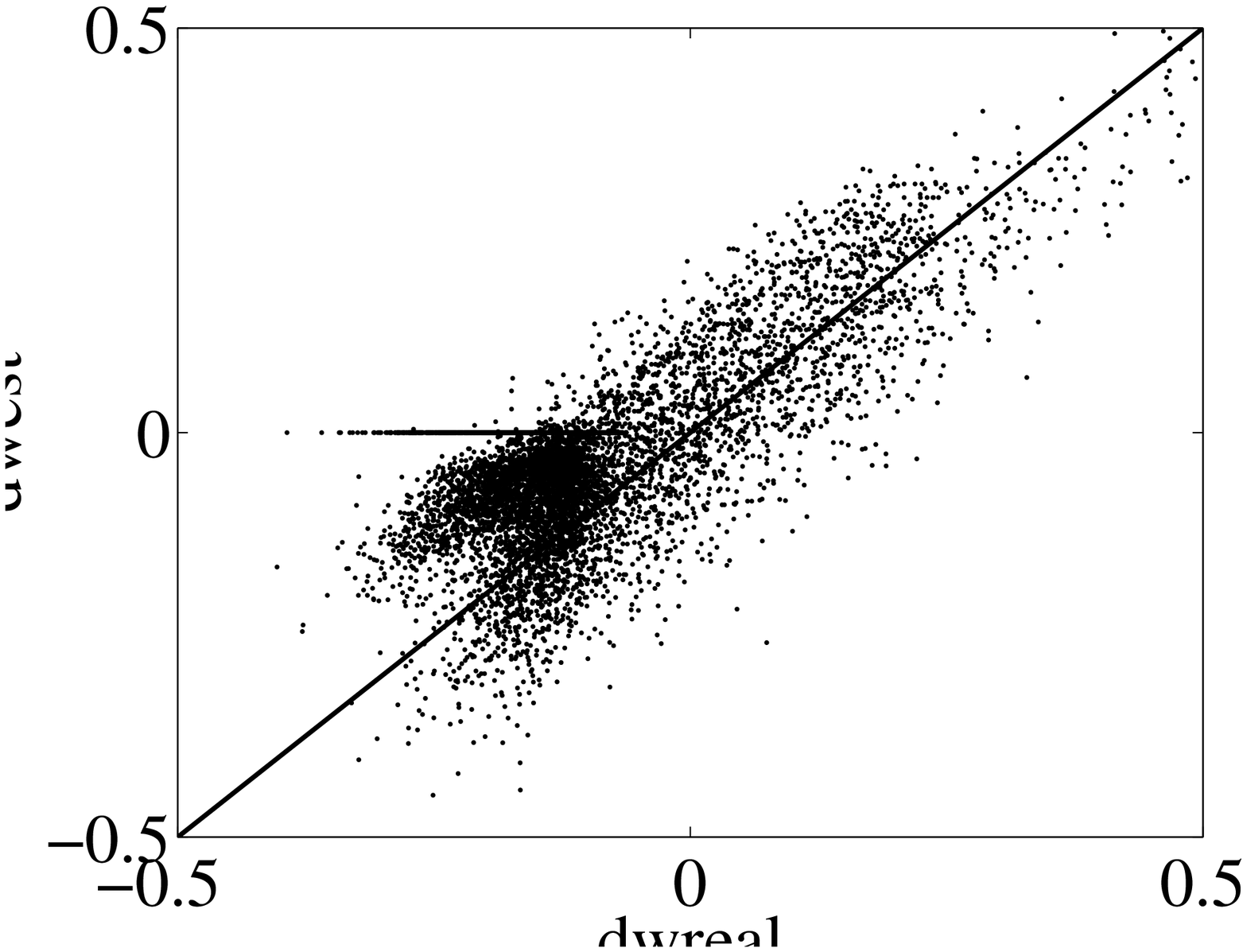}
& \includegraphics[width = 0.3\textwidth]{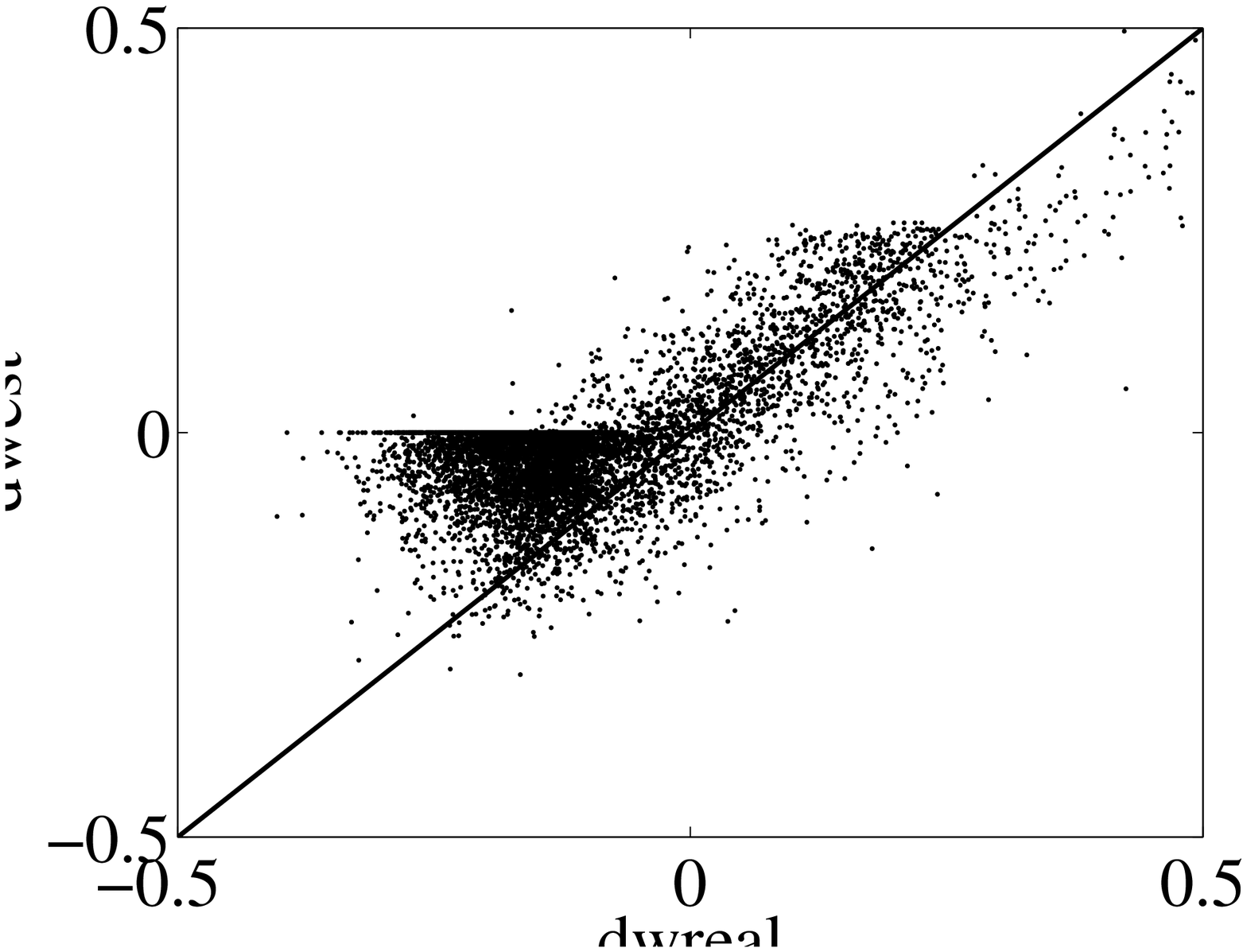}
& \includegraphics[width = 0.3\textwidth]{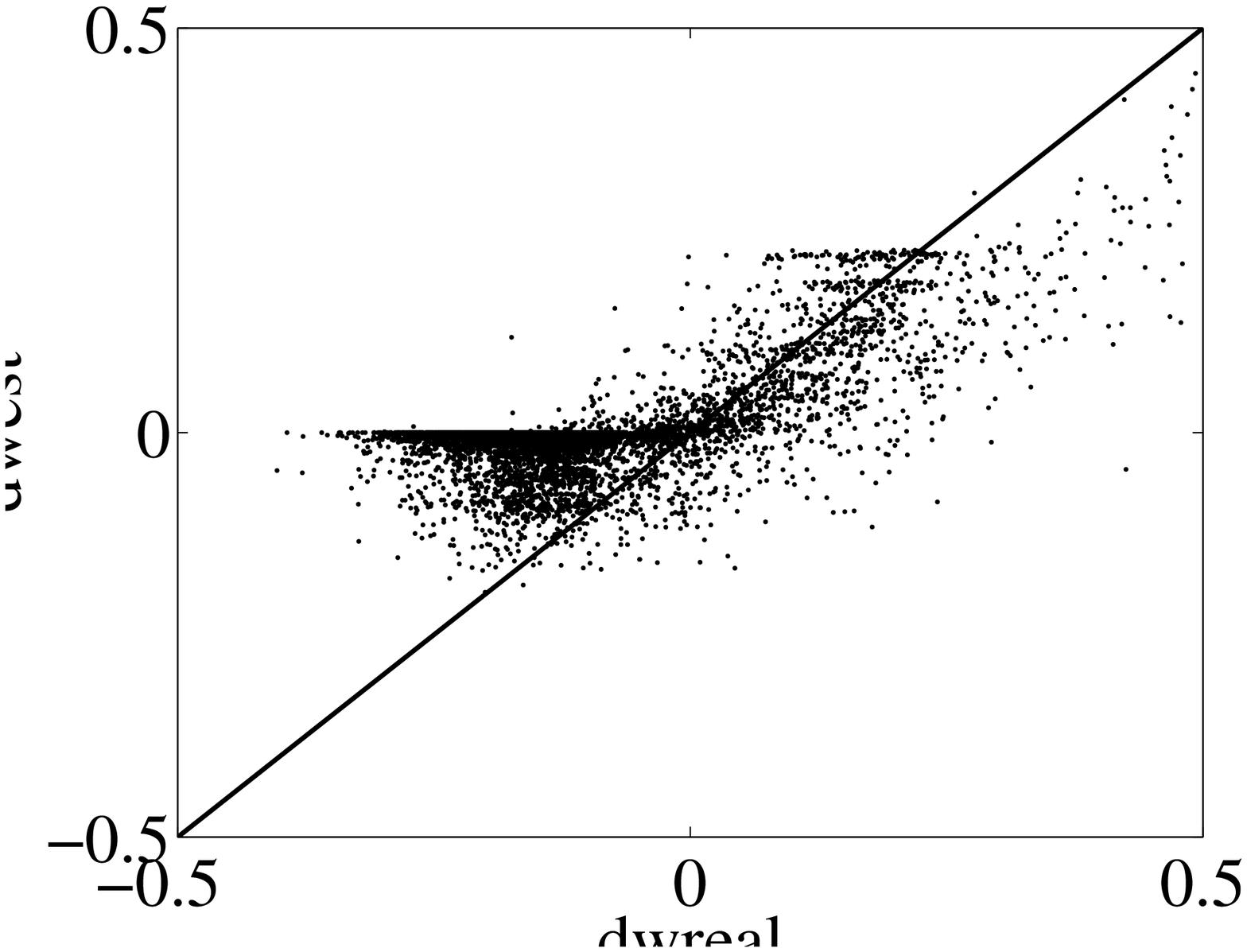}\\
&&\\
\end{tabular}
\caption{Correlation plot between the optimal synaptic weight change $\Delta {\bf w}^{\rm opt} = \Delta {\bf w}^{\Cu}$ and the reconstructed weight change $\Delta {\bf w}^{\rm rec} = C\Delta {\bf W}^{\Cc}$ using the temporal locality constraint. {\bf A}. No locality constraint, i.e. $a = 0$. Deviations from the diagonal are due to the fact that the optimal weight change given by Eq.~(\ref{eq:dwC0}) can not be perfectly accounted for the sum of pair effects. The mean deviations are given by Eq.~(\ref{eq:D}). {\bf B}. A weak locality constraint ($a = 0.04$) almost does not change the quality of the weight change reconstruction. {\bf C}. Strong locality constraint ($a = 0.4$). The horizontal lines arise since most synapses are subject to a few strong updates induced by pairs of pre- and postsynaptic spike times with small time shifts.}
\label{fig:reconstruct}
\end{center}
\end{figure}

The STDP function of Fig.~\ref{fig:C1}A has been derived using a matrix inversion and an ad-hoc locality constraint. In order to assess th errors induced by our approach, we simulate an update $\Delta w^{\rm rec}_{ij} = \sum_{\tpre\in x_j^i}\sum_{\tpost\in y^i}\Delta W^{\Cc}(\tpre-\tpost)$ with the STDP function $\Delta W^{\Cc}$ and compare this update with the exact optimal result $\Delta w_{ij}^{\Cn}$ from Eq.~(\ref{eq:dwC0}).  Fig.~\ref{fig:reconstruct} shows how well the reconstructed weight update $\Delta {\bf w}^{\rm rec} = C\Delta {\bf W}^{\Cc}$ describes the effective weight change $\Delta {\bf w}^{\Cn}$ given by Eq.~(\ref{eq:dwC0}) for various values of the locality parameter $a$. 

\begin{figure}[!h]
\psfrag{dw}{$\Delta w^{\Cc}$}
\psfrag{dt}[tc]{$\tpre-\tpost$}
\begin{center}     
\begin{tabular}{ll}
{\bf A} & {\bf B}\\
\includegraphics[width = 0.4\textwidth]{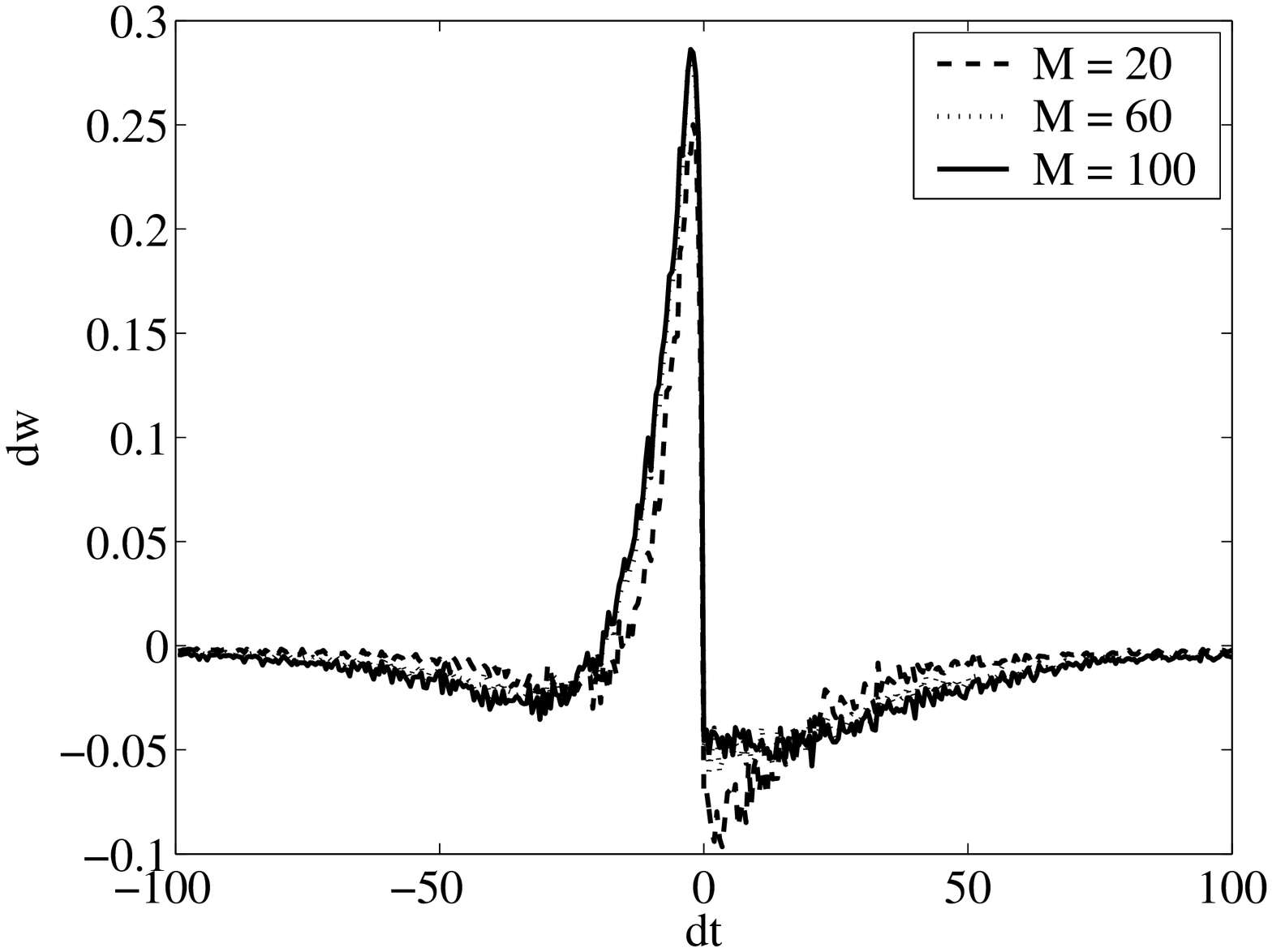}&
\includegraphics[width = 0.4\textwidth]{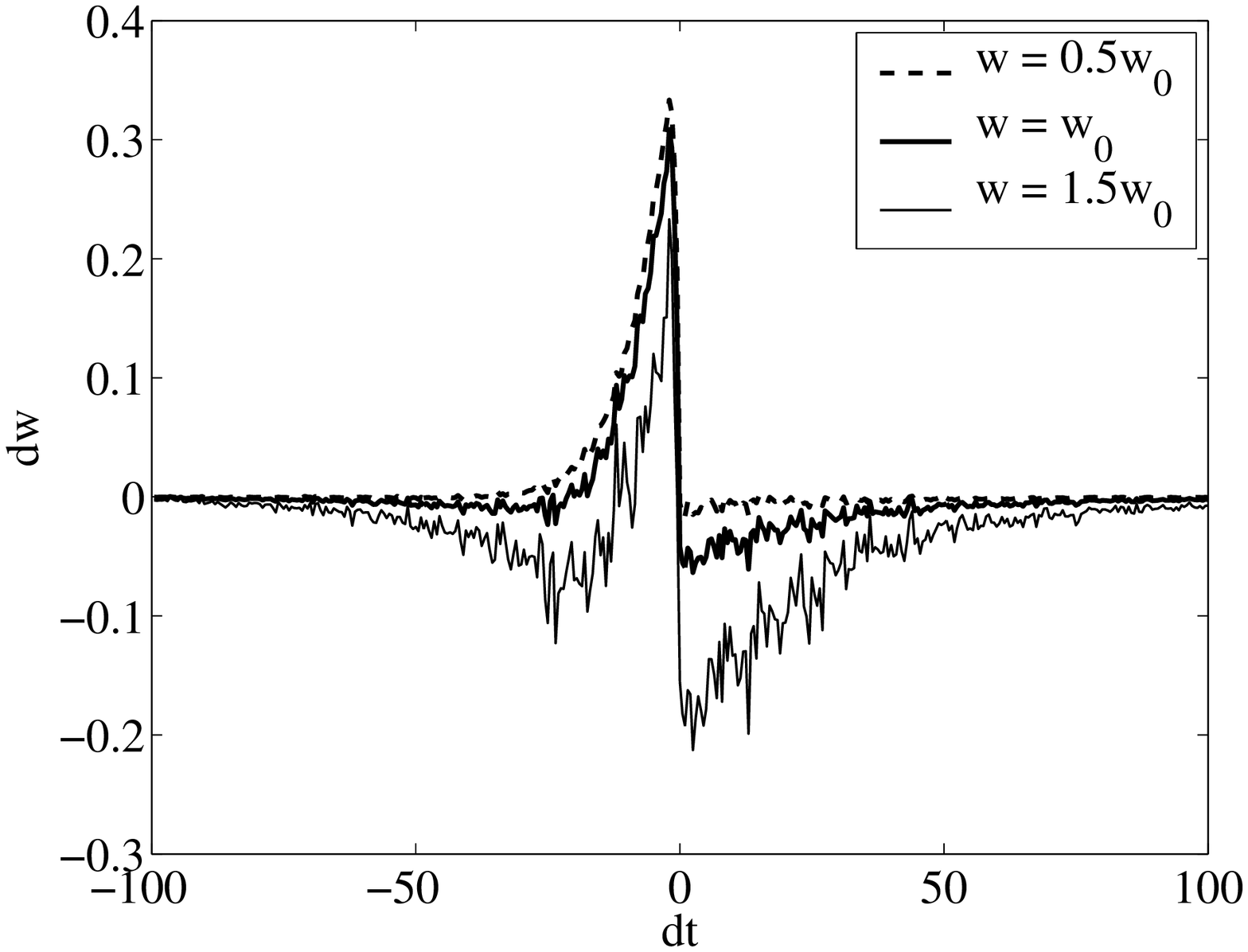}\\
\end{tabular}
\caption{{\bf A}. Optimal STDP function as a function of the number of input patterns $M$. ($a = 0.04$, $N = 400$) {\bf B}. Optimal weight change as a function of the the weight $w$. If the weights are small \emph{(dashed line)} potentiation dominates whereas if they are small \emph{(dotted line)} depression dominates.}
\label{fig:CNSnupre}
\end{center}
\end{figure}

Fig.~\ref{fig:CNSnupre}A shows the STDP functions for various number of patterns $M$. No significant change can be observed for different numbers of input patterns $M$. This is due to the appropriately chosen normalization factor $1/(M-1)$ in the exponent of Eq.~(\ref{eq:LC0}).

The target spike trains $y^i$ have a certain number of spikes during the time window $T$, i.e. they set a target value for the mean rate. Let $\nu^{\rm post} = \frac{1}{TM}\sum_{i=1}^{M}\int_0^Ty^i(t)dt$ be the imposed firing rate. Let $w_0$ denote the amplitude of the synaptic strength such that the firing rate $\bar{\rho}_{w_0}$ given by those weights is identical to the imposed firing rate: $\bar{\rho}_{w_0} = \nu^{\rm post}$. If the actual weights are smaller than $w_0$, almost all the weights should increase whereas if they are bigger than $w_0$, depression should dominate (c.f Fig~\ref{fig:CNSnupre}B). Thus the exact form of the optimal STDP function depends on the initial weight value $w_0$. Alternatively, homeostatic process could assure that the mean weight value is always in the appropriate regime.

\section{Discussion}

\begin{table}[h]
\begin{center}
\begin{tabular}{|l|l||l|l|}
\hline
\multicolumn{2}{|c||}{Unconstrained scenarios} & \multicolumn{2}{c|}{Constrained scenarios}\\
\hhline{|==#==|}
\multirow{2}{3mm}{$\Au$} & \emph{pre-before-post} & \multirow{2}{3mm}{$\Ac$} & \emph{post-before-pre}\\
& LTP $\sim$ EPSP && LTD (or LTP) $\sim$ spike afterpot. \\
\hline
\multirow{2}{3mm}{$\Bu$} &  \emph{pre-before-post} & \multirow{3}{3mm}{$\Bc$} & \emph{post-before-pre}\\
& LTP/LTD $\sim$ reverse correlation && LTD $\sim$ increased firing rate \\
\hline
\multirow{3}{3mm}{$\Cu$} & \emph{pre-before-post} & \multirow{3}{3mm}{$\Cc$} & \emph{post-before-pre} \\
& LTP $\sim$ EPSP & & LTD $\sim$ background patterns \\
& LTD $\sim$ background patterns & & \phantom{LTD} $\sim$ temporal locality\\
\hline
\end{tabular}
\end{center}
\caption{Main results for each scenario.}
\label{tab:res}
\end{table}

Our approach is based upon the maximization of the probability of firing at desired times $\tdes$
with or without constraints.
From the point of view of machine learning,
this is a supervised learning paradigm implemented
as a maximum likelihood approach using the
spike response model with escape noise as a generative model.
Our work   can  be seen as a continuous-time extension of the maximum likelihood approach  proposed by Barber \cite{Barber02b}. 
In the following our discussion will mainly focus
on the relation to  neuroscience (see table~\ref{tab:res}).

From the point of view of neuroscience, paradigms
of unsupervised or reinforcement are probably much
more relevant than the supervised scenario discussed
here. For example, the gradient ascent approach,
used in Scenarios A and B, would yield a slightly
different STDP function, depending on the initialization
of the current set of synaptic weights - and it is not obvious
how one synapse should know about the values of all
other synapses.  Moreover,
it is not clear what a target spike train imposed
by a supervisor really means. One possible interpretation
is that the target spike train is in fact not
necessarily given by a supervisor, but created
spontaneously by the neuron itself. The learning rule
will then change the weights so that the same output
will be more likely when the same stimulus is repeated.
Conditioning the weight changes on the presence
of the a reward signal, would not change
the main results of this paper,
i.e., predictions for the optimal time course
of the STDP function. 

In  scenario $A$, we have mainly shown that the shape of potentiation
in the STDP function is  directly related to  the time course of an EPSP.
Moreover, we have seen that  the behavior of the \emph{post-before-pre} region is determined by the spike afterpotential (see table~\ref{tab:res} for a result summary of the three models). 
In the presence of a teaching input and firing rate constraints, 
a weak reset of the 
membrane potential after the spike 
means that the neuron effectively has a depolarizing spike after potential
(DAP).
In experiments, DAPs have been observed \cite{Feldman00,Markram97,Bi98} for strong presynaptic input.  Other studies,
however,  have shown that the level of depression does not depend on the postsynaptic membrane potential \cite{Sjostrom01}.
In any case, a weak reset (i.e., to a value below threshold
rather than to the resting potential) is consistent with 
the findings of other researchers that used
integrate-and-fire models to account for the high
coefficient of variation of spike trains in vivo
\cite{Bugmann97,Troyer97}.

In the presence of spontaneous activity (c.f. scenario $B$), we have shown that the \emph{pre-before-post} regime  is sensitive to the level of noise and reflects the correlation between pre- and postsynaptic firing. 
Similarly to the observations made for scenario $A$, a constraint on
the spontaneous firing rate causes the optimal weight change
to elicit a  depression of presynaptic spikes that
arrive immediately after the postsynaptic one.
In fact, the reason of the presence of the depression in scenario $\Bc$ is directly related to the presence of a DAP caused by the
strong teaching stimulus. 
In both scenarios $A$ and $B$, depression occurs in order to compensate the increased firing probability due to the DAP.

In scenario $C$, it has been shown that the best way to adapt the weights so that the postsynaptic neurons can detect a specific input patterns among others can be described as a STDP function. This task is similar to the one in \cite{Izhikevich03} in the sense that a neuron is designed to be sensitive to a specific input pattern, but different since our work does not assume any axonal delays.

In our framework, the definition of the objective function is closely related to the neuronal coding. In scenario $C$, we postulate that neurons emit a precise spike train whenever the ``correct'' input is presented and be silent otherwise. This coding scheme is clearly not the most efficient one. Another possibility is to impose postsynaptic neurons to produce a specific but different spike train for each input pattern and not only for the ``correct'' input. Such a modification of the scenario does not dramatically change the results. The only effect is to reduce the amount of depression and increase the amount of potentiation.

Theoretical approaches to neurophysiological phenomena
in general, and to  synaptic plasticity in particular,
can be roughly grouped into three categories:
first, biophysical models that aim at explaining
the STDP function from principles of ion channel
dynamics and intracelluar processes
\cite{Senn01,Shouval02,Abarbanel02,Karmarkar02};
second, mathematical models that start from 
a given STDP function and analyze computational
principles such as intrinsic normalization of summed
efficacies or sensitivity to correlations in the input
\cite{Kempter99c,Roberts99,Roberts00,Rossum00,Kistler00,Song00,Song01,Kempter01,Guetig03};
finally, models that 
derive `optimal'  STDP properties for a given 
computational task \cite{Chechik03,Dayan04,Hopfield04,Bohte04,Bell04,Toyoizumi04}.
Optimizing the likelihood of postsynaptic firing
in a predefined interval, as we did in this paper, is only one possibility
amongst others of introducing concepts
of optimality 
\cite{Barlow61,Atick90,Bell95}
into the field of STDP.
Chechik uses concepts from information
theory \cite{Chechik03}, but restricts his study  to the classification
of stationary patterns. Bohte et al., Bell et al. and Toyoizumi et al. \cite{Bohte04,Bell04,Toyoizumi04} are also using concepts from information theory, but they are applying them to the pre- and postsynaptic spike trains. Dayan use concepts of optimal
filter theory \cite{Dayan04}, but are not interested in precise
firing of the postsynaptic neuron.
The work of Hopfield \cite{Hopfield04} 
is similar to our approach in that it focuses on 
recognition of temporal input patterns, but 
in our approach we are  in addition interested
in triggering postsynaptic  firing 
with precise timing. 
Hopfield and Brody emphasize the 
repair of disrupted  synapses in 
a network that has previously acquired its function
of temporal pattern detector.

Optimality approaches, such as ours, will never be able
to make strict predictions about the real properties of the neuron.
Optimality criteria may, however, help to elucidate
computational principles and provide insights
about potential tasks of electrophysiological phenomena
such as STDP. 


\section*{Acknowledgments}

This work was supported by the Swiss National Science Foundation (200021-100215/1). T.T was supported by the Research Fellowships of the Japan Society for the Promotion of Science for Young Scientists and a Grant-in-Aid for JSPS Fellows. 
\appendix
\section{Probability Density of a Spike Train}
\label{sec:PS}

The probability density of generating a spike train $y_t = \tiens$ with the stochastic process defined by Eq.~(\ref{eq:rho1}) can be expressed as follows:
\begin{equation}
P(y_t) = P(\tio,\dots,\tiF)R(t\vert y_t)
\label{eq:PSt}
\end{equation}

where $P(\tio,\dots,\tiF)$ is the probability density of having $\fmax$ spikes at times $\tio,\dots,\tiF$ and $R(t\vert y_t) = \exp\left(-\int_{\tiF}^{t}\rho(t'\vert S_{t'})dt'\right)$ corresponds to the probability of having no spikes from $\tiF$ to $t$. Since the joint probability $P(\tio,\dots,\tiF)$ can be expressed as a product of conditional probabilities
\begin{equation}
P(\tio,\dots,\tiF) =  P(\tio)\prod_{f=2}^{\fmax}P(\tif\vert \tifm,\dots,\tio)
\end{equation}

Eq. (\ref{eq:PSt}) becomes
\begin{eqnarray}
P(y_t) &=& \rho(\tio\vert y_{\tio})\exp\left(-\int_0^{\tio}\rho(t'\vert y_{t'})dt'\right) \nonumber \\
&\cdot& \left\{\prod_{f=2}^{\fmax}\rho(\tif\vert y_{\tif})\exp\left(-\int_{\tifm}^{\tif}\rho(t'\vert y_{t'})dt'\right)\right\}  \exp\left(-\int_{\tiF}^{t}\rho(t'\vert y_{t'})dt'\right)\nonumber \\
&=& \left(\prod_{\tif\in y_t}\rho(\tif\vert y_{\tif})\right)\exp\left(-\int_0^t \rho(t'\vert y_{t'})dt'\right)
\end{eqnarray}

\section{Numerical Evaluation of $\bar{\rho}(t)$}
\label{sec:numerical}

Since it is impossible to numerically evaluate the instantaneous firing rate $\bar{\rho}(t)$ with the analytical expression given by Eq.~(\ref{eq:brho}), we have to do it in a different way. In fact, there are two different ways to evaluate $\bar{\rho}(t)$. Before going into the details, let us first recall that from the law of large numbers, the instantaneous firing rate is equal to the empirical density of spikes at time $t$:
\be
\Ex{\rho(t\vert y_t)}_{y_t} = \Ex{Y(t)}_{Y(t)}
\label{eq:rho-y}
\ee

where $Y(t) = \sum_{\tif\in y_t}\delta(t-\tif)$ is one realization of the postsynaptic spike train. Thus the first and simplest method based on the r.h.s of Eq.~(\ref{eq:rho-y}) is to build a PSTH by counting spikes in small time bins $[t,t+\delta t]$ over, say $K=10'000$ repetitions of an experiment. The second, and more advanced method, consists in evaluating the l.h.s. of Eq.~(\ref{eq:rho-y}) by Monte-Carlo sampling: instead of averaging over all possible spike trains $y_t$, we generate $K = 10'000$ spike trains by repetition of the same stimulus.  A specific spike train $y_t = \tiens$ will automatically appear with appropriate probability given by Eq.~(\ref{eq:Py}). The Monte-Carlo estimation $\tilde{\rho}(t)$ of $\bar{\rho}(t)$ can be written as
\be
\tilde{\rho}(t) = \frac{1}{P}\sum_{m=1}^{P}\rho(t\vert y_t^m)
\label{eq:semi-anal}
\ee

where $y_t^m$ is the $m^{\rm th}$ spike train generated by the stochastic process given by Eq.~(\ref{eq:rho1}). Since we use the analytical expression of $\rho(t\vert y_t^m)$, we will call Eq.~(\ref{eq:semi-anal}) a semi-analytical estimation.  Let us note that the semi-analytical estimation $\tilde{\rho}(t)$ converges more rapidly to the true value $\bar{\rho}(t)$ than the empirical estimation based on th PSTH.

In the limit of a Poisson process, i.e. $\eta_0 = 0$, the semi-analytical estimation $\tilde{\rho}(t)$ given by Eq.~(\ref{eq:semi-anal}) is equal to the analytical expression of Eq.~(\ref{eq:brho}), since the instantaneous firing rate $\rho$ of a Poisson process is independent of the firing history $y_t = \tiens$ of the postsynaptic neuron.

\section{Deconvolution}
\label{sec:deconv}
\subsection{Deconvolution for Spike Pairs}
\label{sec:deconvpair}

With a learning rule such as Eq.~(\ref{eq:dwC0}), we know the optimal weight change $\Delta w_{ij}$ for each synapse, but we still do not know the corresponding STDP function.

Let us first define the correlation function $c_k(\tau)$, $k=N(i-1)+j$ between the presynaptic spike train $X_j^i(t) = \sum_{\tpre\in x_j^i}\delta(t-\tpre)$ and the postsynaptic spike train $Y^i(t) = \sum_{\tpost\in y^i}\delta(t-\tpost)$:
\begin{equation}
c_k(\tau) = \int_0^TX_j^i(s)Y^i(s+\tau)ds, \qquad k=1,\dots,NM
\label{eq:ck}
\end{equation}

where we allow a range $-T_0\leq\tau\leq T_0$, with $T_0\ll T$.
Since the sum of the pair based weight change $\Delta W$ should be equal to the total adaptation of weights $\Delta w_k$, we can write:
\begin{equation}
\int_{-T_0}^{T_0} c_{k}(s)\Delta W(s) ds \stackrel{!}{=}  \Delta w_k \qquad k = 1,\dots,NM
\end{equation}

If we want to express Eq.~(\ref{eq:ck}) in a matrix form, we need to descretize time in small bins $\delta t$ and define the matrix element 
\begin{equation}
C_{k\Delta} = \int_{\Delta \delta t-T_0}^{(\Delta+1)\delta t-T_0}c_k(s)ds
\end{equation}

Now Eq.~(\ref{eq:ck}) becomes
\begin{equation}
\Delta\mathbf{w} \stackrel{!}{=}  C\Delta\mathbf{W}
\label{eq:dwCDWap}
\end{equation}

where $\Delta \mathbf{w}= (\Delta w_{11},\dots,\Delta w_{1N},\Delta w_{21},\dots,\Delta w_{MN})^T$ is the vector containing all the optimal weight change  and $\Delta \mathbf{W}$ is the vector containing the discretized STDP function, i.e. $\Delta W_{\Delta} = \Delta W(-T_0+\Delta\delta t)$, for $\Delta = 1,\dots,2\tilde{T}_0$ with $\tilde{T}_0 = T_0/\delta t$.

In order to solve the last matrix equation, we have to compute the inverse of the non-square $NM\times 2\tilde{T}_0$ matrix $C$, which is known as the Moore-Penrose inverse (or the pseudo-inverse):
\begin{equation}
C^{+} = (C^TC)^{-1}C^T
\end{equation}

which exists only if $(C^TC)^{-1}$ exists. In fact, the solution given by 
\begin{equation}
\Delta\mathbf{W} = C^{+} \Delta\mathbf{w}
\end{equation}

minimizes the square distance
\begin{equation}
D=\frac{1}{2}(C\mathbf{\Delta W}-\mathbf{\Delta w})^2
\label{eq:D}
\end{equation}

\subsection{Temporal Locality Constraint}
\label{sec:deconvlocal}
If we want to impose a constraint of locality, we can add a term in the minimization process of Eq.~(\ref{eq:D}) and define the following:
\begin{equation}
E=D+\frac{1}{2} \Delta \mathbf{W}^T P \Delta \mathbf{W}
\label{eq:E}
\end{equation}

where $P$ is a diagonal matrix which penalizes non-local terms. In this paper, we take a quadratic suppression of terms that are non-local in time:
\begin{equation}
P_{\Delta\Delta'} = a\delta_{\Delta\Delta'}\left(\Delta-\tilde{T}_0\right)^2
\end{equation}

$\tilde{T}_0$ corresponds to the index of the vector $\Delta {\bf W}$ in Eqs.~(\ref{eq:dwCDWap}) and (\ref{eq:E}) for which $\tpre-\tpost = 0$. Calculating the gradient of $E$ given by Eq.~(\ref{eq:E}) with respect to $\Delta {\bf w}$ yields
\begin{equation}
\nabla_{\Delta \mathbf{W}}E = C^T(C\Delta \mathbf{W}-\Delta \mathbf{w}) + P\Delta\mathbf{W}
\end{equation}

By looking at the minimal value of $E$, i.e. $\nabla_{\Delta W}E = 0$, we have
\begin{equation}
\Delta \mathbf{W} = (C^TC + P)^{-1}C^T\Delta \mathbf{w}
\end{equation}

By setting $a = 0$, we recover the previous case. 







\bibliography{/home/jpfister/doc/paper/refnew_jpp2}
\bibliographystyle{plain}

\end{document}